\newif\ifarxiv
\arxivfalse 

\ifarxiv
\documentclass{jfm-arxiv}
\else
\documentclass{jfm}
\fi
\usepackage{graphicx}
\usepackage{newtxtext}
\usepackage{newtxmath}
\usepackage{natbib}
\usepackage{hyperref}
\hypersetup{
    colorlinks = true,
    urlcolor   = blue,
    linkcolor  = blue,
    citecolor  = black,
}

\newcommand{\RomanNumeralCaps}[1]
\linenumbers

\usepackage{amsmath}
\usepackage{longtable}
\usepackage{enumitem}
\usepackage{siunitx}
\usepackage{physics}
\usepackage{caption}
\usepackage{graphicx}
\usepackage{bm}
\usepackage{float}
\usepackage[dvipsnames]{xcolor}

\definecolor{M030}{HTML}{003554}
\definecolor{M228}{HTML}{006494}
\definecolor{M300}{HTML}{0582ca}
\definecolor{M400}{HTML}{00a6fb}

\def\mytitle{Intrinsic compressibility effects in near-wall turbulence}
\def\myshorttitle{Intrinsic compressibility effects in near-wall turbulence}
\ifarxiv
\title[\myshorttitle]{\vspace{-85pt}\mytitle}
\else
\title[\myshorttitle]{\mytitle}
\fi

\shortauthor{A.~M.~Hasan, P.~Costa, J.~Larsson, S.~Pirozzoli and R.~Pecnik}

\author{Asif~Manzoor~Hasan\aff{1}
  \corresp{\email{a.m.hasan@tudelft.nl}}, 
  Pedro~Costa\aff{1},
  Johan~Larsson\aff{2}, 
  Sergio~Pirozzoli\aff{3} \and
  Rene~Pecnik\aff{1}
 \corresp{\email{r.pecnik@tudelft.nl}}
 }

\affiliation{\aff{1} Process \& Energy Department, Delft University of Technology, Leeghwaterstraat 39, 2628~CB, Delft, The Netherlands
\aff{2}Department of Mechanical Engineering,
University of Maryland,
College Park, MD~20742, USA
\aff{3} Dipartimento
 di Ingegneria Meccanica e Aerospaziale, 
 Sapienza Università di Roma, 
 Via Eudossiana 18,
00184 Roma, Italy}

\begin{document}

\newcommand{\lineMa}{\textcolor{M030}{\rule[1mm]{1.2cm}{0.4mm}}}
\newcommand{\lineMb}{\textcolor{M228}{\rule[1mm]{1.2cm}{0.4mm}}}
\newcommand{\lineMc}{\textcolor{M300}{\rule[1mm]{1.2cm}{0.4mm}}}
\newcommand{\lineMd}{\textcolor{M400}{\rule[1mm]{1.2cm}{0.4mm}}}

\newcommand{\changed}[1]{{\color{red}#1}}

\maketitle

\begin{abstract}

The impact of intrinsic compressibility effects\textemdash changes in fluid volume due to pressure variations\textemdash on high-speed wall-bounded turbulence has often been overlooked or incorrectly attributed to mean property variations. 
To unambiguously quantify these intrinsic compressibility effects, we perform direct numerical simulations of compressible turbulent channel flows with nearly uniform mean properties. Our simulations reveal that intrinsic compressibility effects yield a significant upward shift in the logarithmic mean velocity profile that can be attributed to the reduction in the turbulent shear stress. This reduction stems from the weakening of the near-wall quasi-streamwise vortices. We in turn attribute this weakening to the spontaneous opposition of sweeps and ejections from the near-wall expansions and contractions of the fluid, and provide a theoretical explanation for this mechanism. 
Our results also demonstrate that intrinsic compressibility effects are responsible for the increase in the inner-scaled streamwise turbulence intensity in compressible flows compared to incompressible flows, previously regarded to be an effect of mean property variations.

\end{abstract}

\section{Introduction}

Understanding the impact of compressibility effects on turbulent flow is crucial for a wide range of engineering applications, as it influences the performance and efficiency of aerospace vehicles, gas turbines, combustion processes, and high-speed propulsion systems.
Turbulence in compressible flow involves effects related to heat transfer\textemdash also termed as variable-property effects\textemdash and intrinsic compressibility (hereby IC) effects\textemdash
also termed as `true' compressibility effects \citep{smits2006turbulent}, `genuine' compressibility effects \citep{yu2019genuine}, or simply `compressibility' effects \citep{lele1994compressibility}.
Heat transfer is in turn responsible for two main effects. 
First, heat transfer is associated with mean temperature variations and hence variations in the mean density and viscosity.
Second, it can cause fluctuations in fluid volume (or density) as a result of a change in entropy \citep{livescu2020turbulence}. 
On the other hand, intrinsic compressibility effects are associated with 
changes in fluid volume in response to changes in pressure \citep{lele1994compressibility}. 
While variable-property effects can be relevant at any (even zero) Mach number, IC effects only become important at high Mach numbers.

In 1962, Morkovin postulated that the changes in fluid volume due to entropy and pressure, mentioned above, are negligible such that only mean property variations are important. This hypothesis is commonly referred to as `Morkovin's hypothesis'~\citep{morkovin1962effects,bradshaw1977compressible,coleman1995numerical,smits2006turbulent}. 
Some years later, \cite{bradshaw1977compressible} performed a detailed study on this hypothesis and provided an engineering estimate as to when the hypothesis should hold. According to Bradshaw, Morkovin's postulate may be true in flows where the root-mean-square ($rms$) of the density fluctuation is below 10\% of the mean density.
Subsequently, \cite{coleman1995numerical} noted that most of these density fluctuations arise from passive mixing across mean density gradients. Since Morkovin's hypothesis implicitly assumes that the spatial gradients of the mean density, and thus the fluctuations resulting from them, are small, they argued that the density $rms$ is not a rigorous evaluator of the hypothesis. 
Instead, they claimed that, consistent with the original conjecture, the $rms$ of pressure and total temperature\footnote{We note that pressure fluctuations scaled by mean pressure are a direct measure of intrinsic compressibility effects. However, the justification for why total temperature fluctuations should be small for the hypothesis to hold is unclear \citep{lele1994compressibility}.} scaled by their respective means should be considered. To our knowledge, there is no engineering estimate for these fluctuations such as the one for density proposed by Bradshaw.

If Morkovin's hypothesis holds, then turbulence statistics in compressible flows can be collapsed onto their incompressible counterparts by simply accounting for mean property variations.  
The first key contribution in accounting for variable-property effects was proposed by \cite{van1951turbulent}, who incorporated mean density variations in the mean shear formulation such that 
\begin{equation}\label{vdshear}
\frac{d \bar u}{dy} = \frac{\sqrt{\tau_w/\bar\rho}}{\kappa y}\mathrm{,}   
\end{equation}
where $u$ is the streamwise velocity, $\tau_w$ the wall shear stress, $\rho$ the fluid density, and $\kappa$ the von K\'arm\'an constant. The overbar denotes Reynolds averaging and the subscript $w$ indicates wall values. Equation~\eqref{vdshear} led to two major outcomes: (1) the Van Driest mean velocity transformation \citep{van1956turbulent,danberg1964characteristics} given as 
\begin{equation}\label{vdtrans}
    \bar U_{VD}^+ = \int_0^{\bar u^+} {\sqrt{\frac{\bar \rho}{\rho_w}}} d\bar u^+,
\end{equation}
where the supercript $+$ denotes wall scaling, and (2) the Van Driest skin-friction theory \citep{van1956problem}. These scaling breakthroughs are still widely used, despite their known shortcomings \citep{bradshaw1977compressible,huang1994van,trettel2016mean,patel2016influence,griffin2021velocity,kumar2022modular,hasan2024estimating}. 

Another key contribution is attributed to \citet{morkovin1962effects} who proposed scaling the turbulent shear stress with $\bar \rho /\rho_w$ such that 
\begin{equation}\label{mork}
\widetilde{u^{\prime\prime}v^{\prime\prime}}^* = \frac{\bar \rho}{\rho_w} \frac{\widetilde{u^{\prime\prime}v^{\prime\prime}}}{u_\tau^2}
\end{equation}
collapses with the incompressible distributions. Here, $u_\tau=\sqrt{\tau_w/\rho_w}$ is the friction velocity scale, the tilde denotes density-weighted (Favre) averaging, and the double primes denote fluctuations from Favre average. 
The contributions of Van Driest and Morkovin can be consolidated by interpreting their corrections as if they were to change the definition of the friction velocity scale from $u_\tau$ to $u_\tau^* = \sqrt{\tau_w/\bar \rho}$ (termed `semi-local' friction velocity scale\footnote{The friction velocity scale is termed `semi-local' instead of `local' because the total shear stress in its definition is still taken at the wall.}), such that equations~\eqref{vdshear},~\eqref{vdtrans}, and \eqref{mork} can be rewritten as
\begin{equation}\label{Eq.vd,mork}
    \begin{aligned}
 \frac{d \bar u}{dy} = \frac{u_\tau^*}{\kappa y}\mathrm{,} & & \bar U_{VD}^+ = \int_0^{\bar u} \frac{1}{u_\tau^*} d\bar u, &&\widetilde{u^{\prime\prime}v^{\prime\prime}}^* = \frac{\widetilde{u^{\prime\prime}v^{\prime\prime}}}{u_\tau^{*2}}\mathrm{.}       
    \end{aligned}
\end{equation}

Similarly, efforts to account for mean density and viscosity variations in the definition of the viscous length scale were made since the 1950s \citep{lobb1955nol}, giving rise to the well-known semi-local wall-normal coordinate $y^* = y/\delta_v^*$ (where $\delta_v^* = \bar \mu /(\bar \rho u_\tau^*)$ is the semi-local viscous length scale). Much later, the companion papers by \cite{huang_coleman_bradshaw_1995} and \cite{coleman1995numerical} performed a comprehensive analysis where they showed that turbulence quantities show a much better collapse when reported as a function of $y^*$ rather than $y^+$. Another major consequence of using the semi-local wall coordinate is reflected in velocity transformations. The semi-local velocity transformation, derived independently by \cite{trettel2016mean} and \cite{patel2016influence}, is an extension to the Van Driest velocity transformation accounting for variations in the semi-local viscous length scale. This transformation (also known as the TL transformation) can be written as
\begin{equation}\label{TL}
    \bar U_{TL}^+ = \int_0^{\bar u^+} \left(1 - \frac{y}{\delta_v^*} \frac{d\delta_v^*}{dy} \right) \underbrace{{\frac{u_\tau}{u_\tau^*}}}_{\sqrt{{\bar \rho}/{\rho_w}}} d\bar u^+.
\end{equation}
In short, the above-mentioned scaling theories in equations~\eqref{Eq.vd,mork}~and~\eqref{TL} show that heat transfer effects associated with mean property variations can be accounted for in terms of the semi-local friction velocity and viscous length scales.

In addition to the studies mentioned above, many other studies have addressed variable-property effects in low-Mach~\citep{patel2015semi} and high-Mach number flows~\citep[][to name a few]{maeder2001direct,morinishi2004direct,foysi2004compressibility,duan2010direct,duan2011direct,modesti2016reynolds,zhang2018direct,cogo2022direct,Zhang_Wan_Liu_Sun_Lu_2022,wenzel2022influences,cogo2023assessment}. 
However, less emphasis has been placed on studying intrinsic compressibility effects, possibly due to the belief that Morkovin's hypothesis holds for wall-bounded flows even in the hypersonic regime~\citep{duan2011direct,zhang2018direct}.

Recently, by isolating intrinsic compressibility effects, \cite{hasan2023incorporating} found that Morkovin's hypothesis is inaccurate at high Mach numbers. These compressibility effects modify the mean velocity scaling, leading to an upward shift in the logarithmic profile. The authors attributed this shift to the modified near-wall damping of turbulence and proposed a mean velocity transformation based on a modification of the Van Driest damping function as
\begin{equation}
    \bar U_{HLPP}^+ = \int_0^{\Bar{u}^+} \! \!
     \left({ \frac{1 + \kappa y^* {D(y^*,M_\tau)}} {1 + \kappa {y^*} {D(y^*,0)}}}\right){\left({1 - \frac{y}{\delta_v^*}\frac{d \delta_v^*}{dy}}\right)} \sqrt{\frac{\Bar{\rho}}{\rho_w}} \, {d \Bar{u}^+}.
     \label{eq:asif}
\end{equation}
This transformation was found to be accurate for a wide variety of flows including (but not limited to) adiabatic and cooled boundary layers, adiabatic and cooled channels, supercritical flows, and flows with non-air-like viscosity laws. 
The modified damping function in \eqref{eq:asif} reads 
\begin{equation}
D(y^*,M_\tau) = \left[1 - \mathrm{exp}\left({\frac{-y^*}{A^+ + f(M_\tau)}}\right)\right]^2,
\end{equation}
with $f(M_\tau) = 19.3M_\tau$.
Despite the evidence that intrinsic compressibility effects modify the damping, the underlying physical mechanism is still unknown. 

More evidence on the importance of intrinsic (or `genuine') compressibility effects has been provided in a series of recent publications by
Yu and co-workers~\citep{yu2019genuine,yu2020compressibility,yu2021compressibility}, who analysed these effects in channel flows through direct numerical simulations (DNS). They performed a Helmholtz decomposition of the velocity field and mainly focused on dilatational motions and their direct contribution to several turbulence statistics. Their main observations were: (1) intrinsic compressibility effects, if present, are likely concentrated in the near-wall region, where the wall-normal dilatational velocity field exceeds the solenoidal counterpart; (2) the correlation between the solenoidal streamwise and the dilatational wall normal velocity is negative and can constitute up to 10\% of the total shear stress; (3) this negative correlation was attributed to the opposition of sweeps near the wall by dilatational motions; and (4) the dilatation field (and thus the dilatational velocity) exhibits a travelling wave-packet-like structure, whose origin is yet unknown \cite[see also][]{tang2020near, gerolymos2023scaling,yu2024generation}.

In this paper, we will focus mainly on the indirect effects of intrinsic compressibility, namely, those that do not result directly from contributions by dilatational motions but result as a consequence of changes in the solenoidal dynamics of turbulence. 
To achieve this, we first perform direct numerical simulations employing the methodology described in \cite{coleman1995numerical}, whereby variable-property effects are essentially removed by cancelling the aerodynamic heating term in the energy equation. These simulations will allow us to study intrinsic compressibility effects by isolating them. 
With this approach, our main goal is to answer why the near-wall damping of turbulence changes with increasing Mach number, as observed in \cite{hasan2023incorporating}. Since this is also observed for conventional flows, we believe that the knowledge obtained from our simplified cases is directly applicable to those flows. 
With the simulated cases, we look into various fundamental statistics of turbulence such as turbulent stresses, pressure-strain correlation, and into coherent structures, eventually tracing back the change in near-wall damping of the turbulent shear stress to the weakening of quasi-streamwise vortices. Subsequently, with the help of what is known from the incompressible turbulence literature, we provide a theoretical explanation as to why the vortices weaken.

The paper is structured as follows. \S\ref{Sec:2} describes the cases and methodology used in this paper. \S\ref{Sec:3} explains the change in damping of near-wall turbulence as a result of the change in turbulent stress anisotropy, caused by a reduction in the pressure-strain correlation.
\S\ref{Sec:4} connects this reduced correlation with the weakening of quasi-streamwise vortices, which is then explained using conditional averaging. Finally, the summary and conclusions are presented in \S\ref{Sec:5}.   

\section{Computational approach and case description}\label{Sec:2}

In order to investigate turbulence in high-speed wall-bounded channel flows with uniform mean temperature (internal energy) in the domain, we perform direct numerical simulations by solving the compressible Navier-Stokes equations in conservative form, given as
\begin{equation}
\begin{aligned}
\frac{\partial \rho}{\partial t}+\frac{\partial \rho u_{i}}{\partial x_{i}} &=0, \\
\frac{\partial \rho u_{i}}{\partial t}+\frac{\partial \rho u_{i} u_{j}}{\partial x_{j}} &=-\frac{\partial p}{\partial x_{i}}+\frac{\partial \tau_{i j}}{\partial x_{j}}+f \delta_{i 1}, \\
\frac{\partial \rho E}{\partial t}+\frac{\partial \rho u_{j} E}{\partial x_{j}} &=-\frac{\partial p u_j}{\partial x_{j}}-\frac{\partial q_{j}}{\partial x_{j}}+\frac{\partial \tau_{i j} u_{i}}{\partial x_{j}}+f u_{1} + \Phi.
\end{aligned}
\end{equation}
The viscous stress tensor and the heat flux vector are given as  
\begin{equation}\label{Eq:tauij_qij}
\tau_{i j}=\mu\left(\frac{\partial u_{i}}{\partial x_{j}}+\frac{\partial u_{j}}{\partial x_{i}}-\frac{2}{3} \frac{\partial u_{k}}{\partial x_{k}} \delta_{i j}\right), ~
q_{j}=-\lambda \frac{\partial T}{\partial x_{j}},
\end{equation}
where $u_i$ is the velocity component in the $i^{th}$
direction, and where $i= 1, 2, 3$ corresponds to the streamwise ($x$), wall-normal ($y$) and spanwise ($z$) directions, respectively. $\rho$ is the density, $p$ the pressure, $E = c_v T + u_i u_i /2$ the total energy per unit mass, $\mu$ the viscosity, $\lambda$ the thermal conductivity and $Pr=\mu c_p/\lambda$ the Prandtl number. $c_p$ and $c_v$ indicate specific heats at constant pressure and constant volume, respectively. $f$ is a uniform body force that is adjusted in time to maintain a constant total mass flux in periodic flows (e.g., a fully developed turbulent channel or pipe). 

As outlined in the introduction, herein we attempt to remove mean property gradients to isolate intrinsic compressibility effects. For that purpose, we follow the approach presented by \cite{coleman1995numerical}, whereby the energy equation is augmented with a source term
\begin{equation}
\Phi = -\tau_{i j}\frac{\partial u_i}{\partial x_{j}}
\end{equation}
that counteracts the effects of viscous dissipation. Consequently, the mean internal energy remains approximately uniform across the entire domain. For an ideal gas, this implies that the mean temperature is also approximately constant, which, when combined with a uniform mean pressure, leads to a nearly uniform mean density. Furthermore, the mean dynamic viscosity and mean thermal conductivity are also uniform. However, it is important to note that the simulations still permit fluctuations of these properties\textemdash primarily along isentropes, as we will see below.

Using this approach, four cases with increasing Mach numbers are simulated, as presented in table~\ref{casetab}. These simulations are performed with \texttt{STREAmS} \citep{bernardini2021streams} using the assumption of a calorically perfect ideal gas (constant specific heat capacities), a constant Prandtl number of $0.7$, and a power law for the viscosity with an exponent of $0.75$. The domain is periodic in the streamwise and spanwise directions, while at the walls an isothermal boundary condition is used for temperature, and a zero normal gradient is specified for pressure. 
Since the four cases have similar $Re_\tau$ values, we use the same grid for all simulations. The computational grid consists of $n_x = 1280$, $n_y = 480$ and $n_z = 384$ points for a domain of size $L_x = 10h$, $L_y = 2h$ and $L_z = 3h$, where $h$ is the channel half-height. This gives a near-wall resolution of $\Delta x^+ = 4.3$ and $\Delta z^+ = 4.3$. The grid in the wall-normal direction is stretched in such a way that $y^+ \leq 1$ is achieved for the first grid point. 

\begin{table}
\centering
\begin{tabular}{m{2.4cm} m{1.5cm} m{1.5cm} m{1.5cm} m{1.5cm} m{1.5cm} m{1.5cm}}
Case name &  $M_{b}$ & $M_{cl}$ &$M_{\tau}$ & $Re_{\tau}$ & $Re_{\tau_c}$   & Line colour \\ \hline
Mach 0.3 & 0.3 & 0.34  & 0.0162 & 556  & 556  & \lineMa \\ 
Mach 2.28 & 2.28 & 2.59 & 0.1185  & 546  & 539 & \lineMb \\ 
Mach 3 & 3.0 & 3.37  & 0.1526  & 547 & 527 & \lineMc \\ 
Mach 4 & 4.0 & 4.47 & 0.1968 & 544 & 513 & \lineMd \\ 
\end{tabular}
\captionof{table}{Description of the cases. $M_{b} = U_b / \sqrt{\gamma R T_w}$ is the bulk Mach number, $M_{cl} = U_c / \sqrt{\gamma R T_c}$ is the channel centreline Mach number and $M_{\tau} = u_\tau / \sqrt{\gamma R T_w}$ is the wall friction Mach number. $Re_\tau = \rho_w u_\tau h/\mu_w$ is the friction Reynolds number based on the channel half-height $h$ and $Re_{\tau_c}$ corresponds to the value of the semi-local friction Reynolds number ($Re_\tau^* = \bar \rho u_\tau^* h/\bar \mu$) at the channel centre.}
\label{casetab}
\end{table}

\begin{figure}
	\centering	\includegraphics[width=\textwidth]{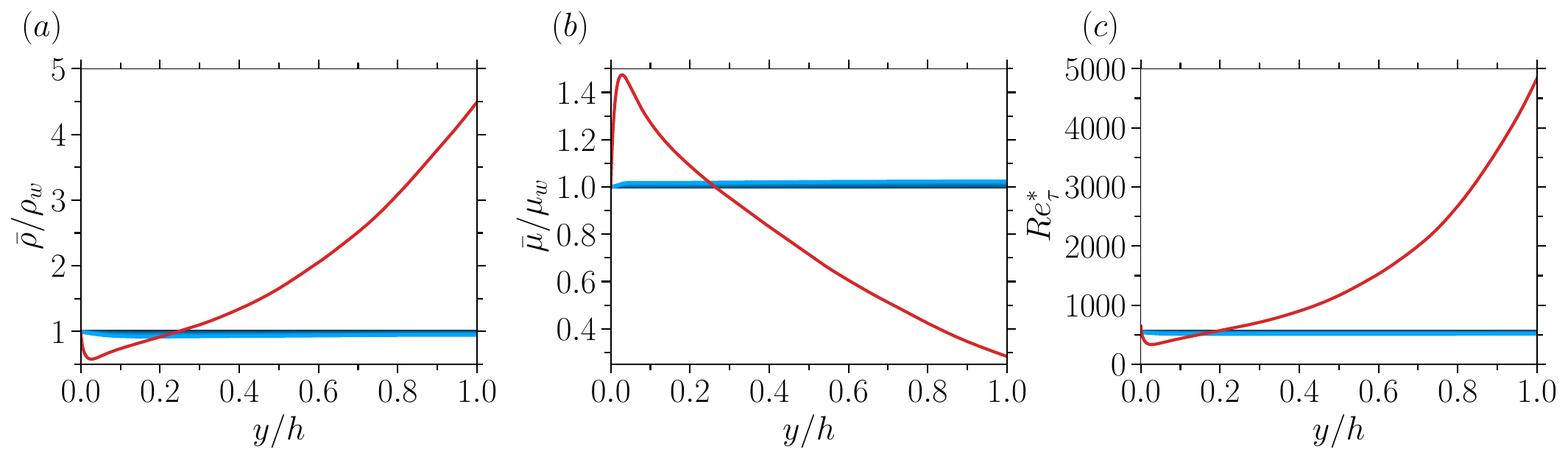}
	\caption{Wall-normal distributions of  (a) density $\overline{\rho}$, (b) viscosity $\overline{\mu}$, and (c) the semi-local friction Reynolds number $Re_\tau^* = \bar \rho u_\tau^* h/\bar \mu$ for the cases described in table~\ref{casetab}. The red lines represent the $M_\infty=14$ case of \cite{zhang2018direct}. These quantities are plotted as a function of the wall-normal coordinate scaled by the channel half-height for the channel flow cases, and by boundary layer thickness ($\delta_{99}$) for the $M_\infty=14$ boundary layer case.}
 \label{Fig:prop}
\end{figure}

Figure~\ref{Fig:prop} shows the mean density, viscosity, and semi-local Reynolds number profiles for the four cases introduced in table~\ref{casetab}. The figure also shows the profiles of a conventional boundary layer at a free-stream Mach number of 14, taken from \cite{zhang2018direct}. 
Compared to the conventional $M_\infty=14$ boundary layer case, our cases show little to no variation in mean properties. This implies that mean heat transfer effects are indeed negligible in the present cases. 

\begin{figure}
    \centering	
    \includegraphics[width=\textwidth]{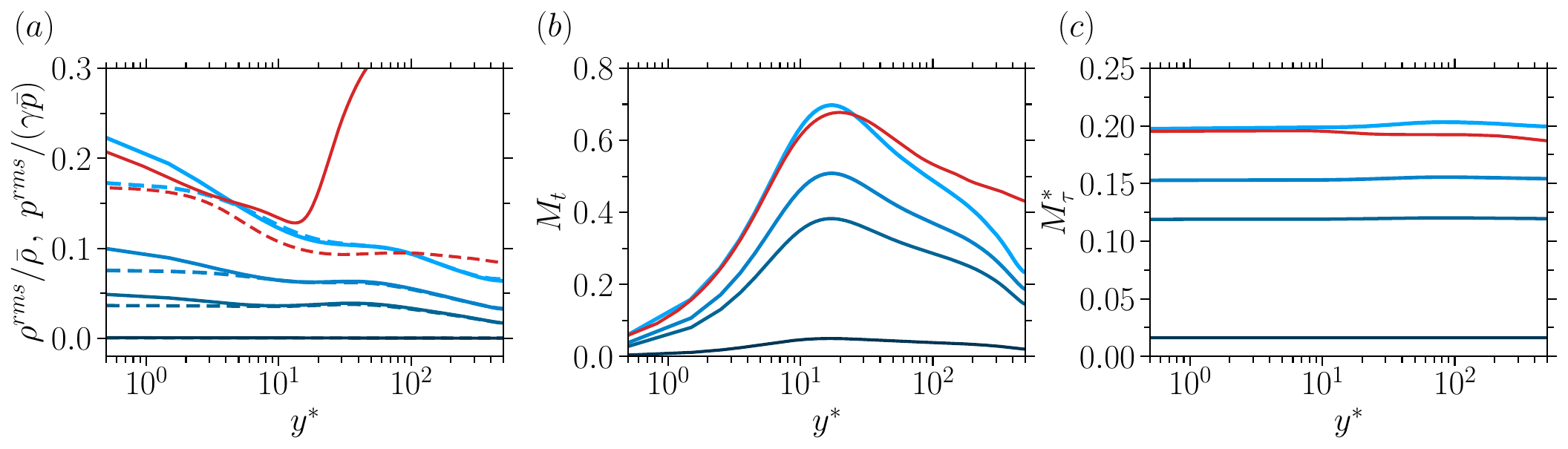}
    \caption{Wall-normal distributions of (a) the root-mean-square ($rms$) of the total (solid) and isentropic (dashed) density fluctuations [equation~\eqref{rhoisen}]; (b) the turbulence Mach number $M_t=\sqrt{2k}/\sqrt{\gamma R \bar T}$; and (c) the semi-local friction Mach number $M_\tau^* = u_\tau^*/\sqrt{\gamma R \bar T}$ for the cases described in table~\ref{casetab}. The red lines represent the $M_\infty=14$ case of \cite{zhang2018direct}.}
\label{Fig:prms}
\end{figure}

To determine whether other heat transfer effects associated with changes in fluid volume as a result of changes in entropy are important, we compute density fluctuations using the isentropic relation 
\begin{equation}\label{rhoisen}
    \frac{\rho^{is}_{rms}}{\bar \rho} \approx \frac{1}{\gamma}\frac{p_{rms}}{\bar p},
\end{equation}
and compare it with the density fluctuations obtained from DNS in figure~\ref{Fig:prms}(a). 
With the exception of the viscous sublayer, the two distributions appear to collapse, which implies that entropic heat transfer effects are negligible in the present cases. Hence, any deviations from incompressible flows observed in these cases should be attributed to intrinsic compressibility effects.

Figure~\ref{Fig:prms}(a) also shows the total and isentropic density fluctuations for the $M_\infty = 14$ flow case computed by \cite{zhang2018direct}.
As can be seen, the total density fluctuations are much higher than the isentropic ones  
in the buffer layer and beyond, 
corroborating that both heat transfer and intrinsic compressibility effects are important. Interestingly, our highest Mach number case (Mach 4) and Zhang's $M_\infty=14$ boundary layer have similar isentropic density $rms$ (or similar pressure $rms$). Given that the pressure $rms$ scaled by mean pressure is an effective measure of intrinsic compressibility effects \citep{coleman1995numerical}, we can expect that these effects are of comparable magnitude for our Mach 4 case and the conventional $M_\infty=14$ boundary layer.

In addition to the pressure $rms$, intrinsic compressibility effects can also be quantified in terms of Mach numbers. Figure~\ref{Fig:prms}(b) shows the turbulence Mach number, defined as $M_t=\sqrt{2k}/\sqrt{\gamma R \bar T}$, where $k = \overline{\rho u_i^{\prime\prime}u_i^{\prime\prime}}/2$ is the turbulence kinetic energy (TKE) and the denominator is the local speed of sound for ideal gases.
Three out of four cases are above the threshold of $M_t=0.3$, above which intrinsic compressibility effects are considered important \citep{smits2006turbulent}. Due to the inhomogeneous nature of wall-bounded flows, $M_t$ is not constant throughout the domain, becoming zero at the wall where the pressure and density $rms$ are the strongest as shown in figure~\ref{Fig:prms}(a). 

Other parameters have been proposed in the literature as a better measure of intrinsic compressibility effects in wall-bounded flows, most prominently the friction Mach number $M_\tau = u_\tau/\sqrt{\gamma R T_w}$ \citep{bradshaw1977compressible, smits2006turbulent,yu2022wall, hasan2023incorporating}. When defined in terms of local properties, one obtains the semi-local friction Mach number $M_\tau^* = u_\tau^*/\sqrt{\gamma R \bar T}$. 
Figure~\ref{Fig:prms}(c) shows that, in contrast to $M_t$, the distribution of $M_\tau^*$ is nearly constant, even for flows with mean property variations. The reason why $M_\tau^*$ is constant for flows with ideal gases is because $\bar T/T_w \approx \rho_w/\bar \rho$ such that 
\begin{equation}\label{Eq:Mtau}
    M_\tau^* = \frac{u_\tau^*}{\sqrt{\gamma R \bar T}} = \frac{u_\tau \sqrt{\rho_w/\bar\rho}}{\sqrt{\gamma R \bar T}}\approx \frac{u_\tau \sqrt{\bar T/T_w}}{\sqrt{\gamma R \bar T}} = \frac{u_\tau}{\sqrt{\gamma R T_w}} = M_\tau.
\end{equation}
As seen in figure~\ref{Fig:prms}(b)~and~(c), the profiles of $M_t$ and $M_\tau^*$ are equivalent for the Mach 4 constant-property and the $M_\infty=14$ conventional cases, further supporting the statement made above that the IC effects in these cases are comparable.

\section{Intrinsic compressibility effects on turbulence statistics}\label{Sec:3}
Having introduced the flow cases, we
first discuss the modified near-wall damping of the turbulent shear stress and its consequence on the mean velocity scaling. 
Unless otherwise stated, all quantities will be presented in their semi-locally scaled form. Nevertheless, since the cases have approximately constant mean properties, there is no major difference between the classical wall scaling (denoted by the superscript~`$+$') and the semi-local scaling (denoted by the superscript~`$*$').

\subsection{Outward shift in viscous and turbulent shear stresses}

In the inner layer of parallel (or quasi-parallel) shear flows, integration of the mean streamwise momentum equation implies that the sum of viscous and turbulent shear stresses is equal to the total shear stress, given as
\begin{equation}\label{eq1}
\overline{\mu\left(\frac{\partial u}{\partial y} + \frac{\partial v}{\partial x}\right)} - \overline{\rho u^{\prime\prime} v^{\prime\prime}} = {\tau_{tot}}
,
\end{equation}
where $\tau_{tot}\approx \tau_w$ in zero-pressure-gradient boundary layers, whereas it decreases linearly with the wall distance in channel flows. 
Neglecting terms due to viscosity fluctuations and
normalizing equation~\eqref{eq1} by $\tau_w$, we get for the latter case
\begin{equation}\label{eq2}
\frac{\bar{\mu}}{\mu_w} \frac{d \bar{u}^+}{d y^+} -\widetilde{u^{\prime\prime} v^{\prime\prime}}^* \approx 1 - \frac{y}{h}
,
\end{equation}
where $h$ is the channel half-height.

\begin{figure}
	\centering
 \includegraphics[width=1\textwidth]{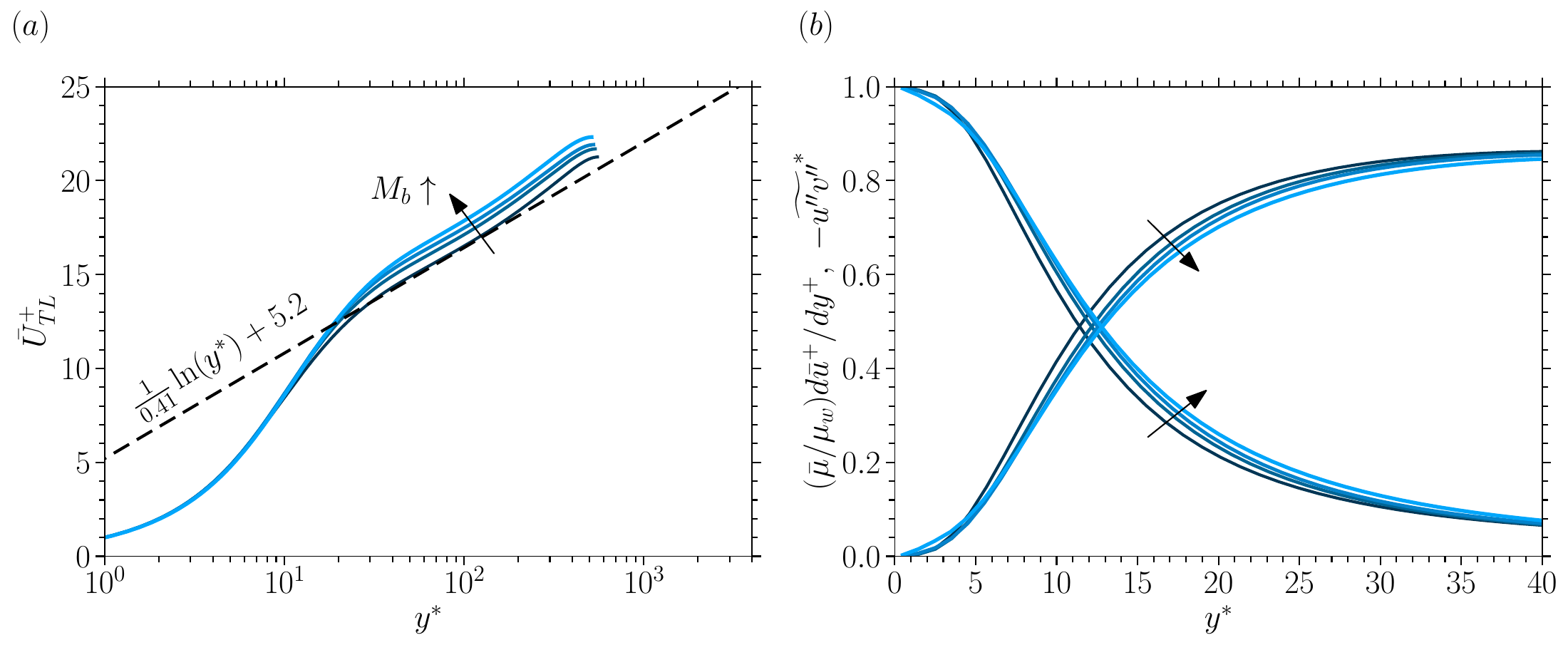}
	\caption{(a) TL-transformed mean velocity profiles [equations~\eqref{TL}, \eqref{utl}], and (b) viscous and turbulent shear stresses for the cases described in table~\ref{casetab}.}
\label{Fig:shearstress}
\end{figure}

Integrating the viscous shear stress yields the TL-transformed mean velocity profile \citep{trettel2016mean,patel2016influence} as
\begin{equation}\label{utl}
    \bar U^+_{TL} = \int_0^{y^*} \frac{\bar{\mu}}{\mu_w} \frac{d \bar{u}^+}{d y^+} dy^*  \mathrm{.}
\end{equation}
Figure~\ref{Fig:shearstress}(a) shows the transformed velocity profiles for the cases listed in table~\ref{casetab} (or simply $\bar u^+$, since the mean flow properties are nearly constant). A clear shift in the logarithmic profile is seen that increases with the Mach number. Based on equation~\eqref{utl}, an upward shift in the mean velocity profile corresponds to an equivalent upward shift (or increase) in the viscous shear stress. This is evident from figure~\ref{Fig:shearstress}(b).
Since the total shear stress is universal for the four flow cases under inspection, an increase in the viscous shear stress directly implies a decrease in the turbulent shear stress. Indeed, figure~\ref{Fig:shearstress}(b) shows that the turbulent shear stress reduces with increasing Mach number.
In other words, the log-law shift observed in figure~\ref{Fig:shearstress}(a) is a consequence of the modified damping of the turbulent shear stress, as also noted by \cite{hasan2023incorporating}. 

\subsection{Outward shift in wall-normal turbulent stress: change in turbulence anisotropy}

The outward shift in the turbulent shear stress corresponds to an outward shift in the wall-normal turbulent stress, because wall-normal motions directly contribute to turbulent shear stress by transporting momentum across the mean shear \citep{townsend1961equilibrium, deshpande2021active}. This is also reflected in the turbulent shear stress budget, whose production is controlled by the wall-normal turbulent stress~\citep{pope2001turbulent}. 

\begin{figure}
	\centering	\includegraphics[width=1\textwidth]{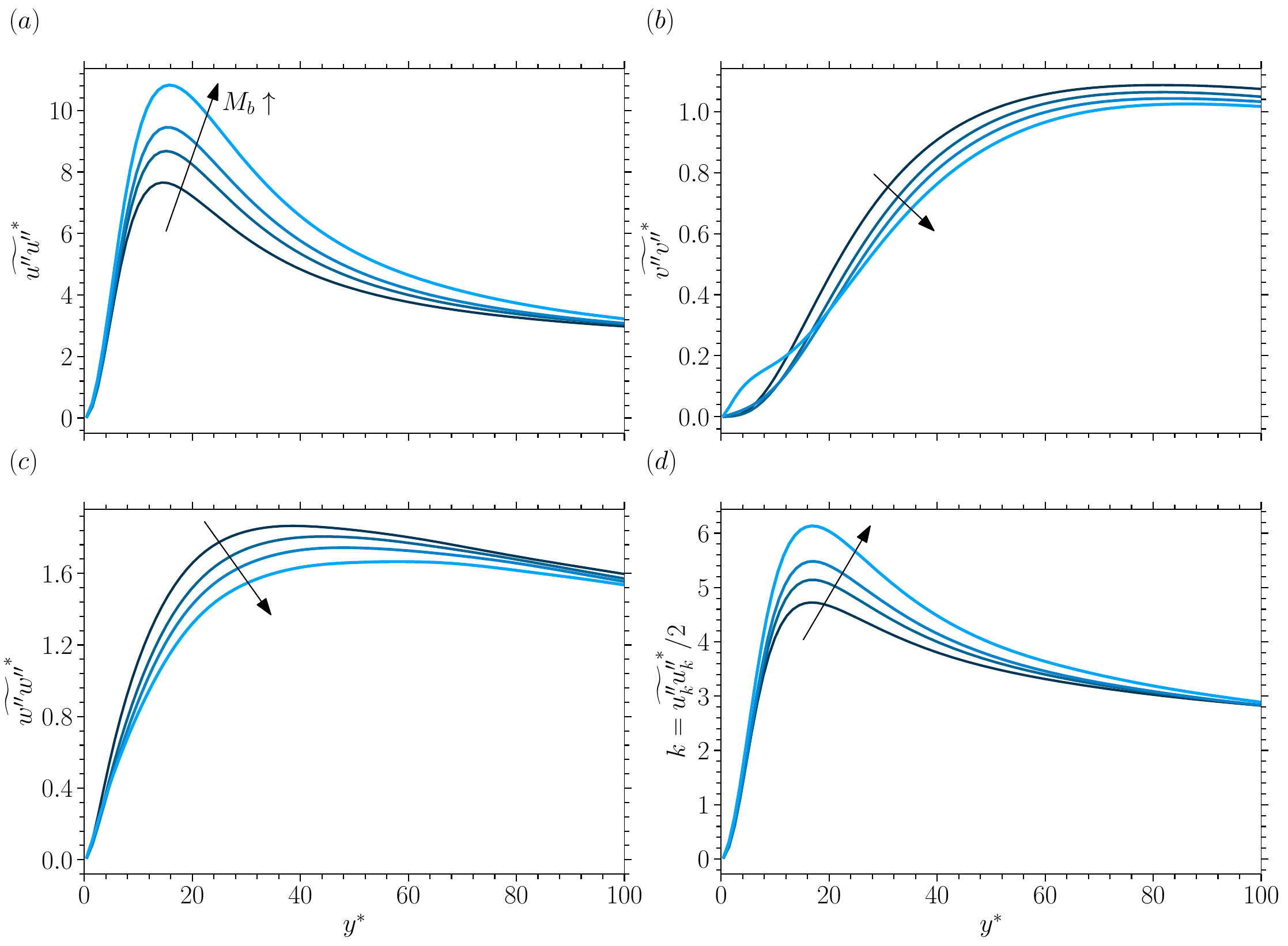}
	\caption{Wall-normal distributions of (a) streamwise, (b) wall-normal and (c) spanwise turbulent stresses, and (d) the turbulence kinetic energy for the cases described in table~\ref{casetab}.}
\label{Fig:normalstress}
\end{figure}
Figure~\ref{Fig:normalstress}(b) shows profiles of the wall-normal turbulent stress. A clear outward shift is evident, which is consistent with the observed outward shift in the turbulent shear stress. Now, the decrease in the wall-normal stress can either be due to less energy being received from the streamwise component (inter-component energy transfer), or due to an overall reduction of the turbulence kinetic energy. 
In order to clarify this, 
we report the streamwise and the spanwise turbulent stresses, along with the turbulence kinetic energy in panels~(a),~(c) and (d) of figure~\ref{Fig:normalstress}, respectively.

Figure~\ref{Fig:normalstress}(a) shows that the streamwise turbulent stress becomes stronger with increasing Mach number. The increase in the peak streamwise turbulence intensity in compressible flows, compared to incompressible flows at similar Reynolds numbers, has also been observed in several other studies 
\citep{gatski2002numerical,pirozzoli2004direct,foysi2004compressibility,duan2010direct,modesti2016reynolds,zhang2018direct,trettel2019transformations,cogo2022direct,cogo2023assessment}. However, none of these studies assessed whether intrinsic compressibility effects play a role in peak strengthening. In fact, the higher peak observed in the $M_\infty = 14$ boundary layer was attributed to variable-property effects by \cite{zhang2018direct}. 
Our results instead demonstrate unambiguously that intrinsic compressibility effects play a central role in the strengthening of streamwise turbulence intensity, since our flow cases are essentially free of variable-property effects.

Similar to the wall-normal stress, the spanwise turbulent stress also decreases with increasing Mach number, shown in figure~\ref{Fig:normalstress}(c). 
The increase in the streamwise stress and the decrease in the wall-normal and spanwise stresses imply suppression of inter-component energy transfer with increasing Mach number. However, before discussing this in more detail in the next subsection, we first note that the increase in the streamwise turbulent stress is much more pronounced than the decrease in the other two components, which essentially results in an increase in the turbulence kinetic energy with Mach number as shown in  figure~\ref{Fig:normalstress}(d). This suggests that, in addition to the change in intercomponent energy transfer, there is also a change in the production of $\widetilde{u^{\prime\prime}u^{\prime\prime}}^*$. This change in production can be attributed to the changes in viscous and turbulent shear stresses observed in figure~\ref{Fig:shearstress}, since it is their product that governs the production term. This is further discussed in detail in Appendix~\ref{Sec:streamwise}, where we present the budget of the streamwise turbulence stress, and provide a phenomenological explanation for the increase in $\widetilde{u^{\prime\prime}u^{\prime\prime}}^*$.

\subsection{Reduced inter-component energy transfer}\label{subsec:intercomp}

The strengthening of the streamwise turbulent stress and the weakening of the other two components, as observed in figures~\ref{Fig:normalstress}(a)~-~(c), imply an increase in turbulence anisotropy, which was also previously observed in several studies on compressible wall-bounded flows \citep{foysi2004compressibility, duan2010direct,zhang2018direct, cogo2022direct,cogo2023assessment}, mainly regarded as a variable-property effect.

\begin{figure}
	\centering	\includegraphics[width=1\textwidth]{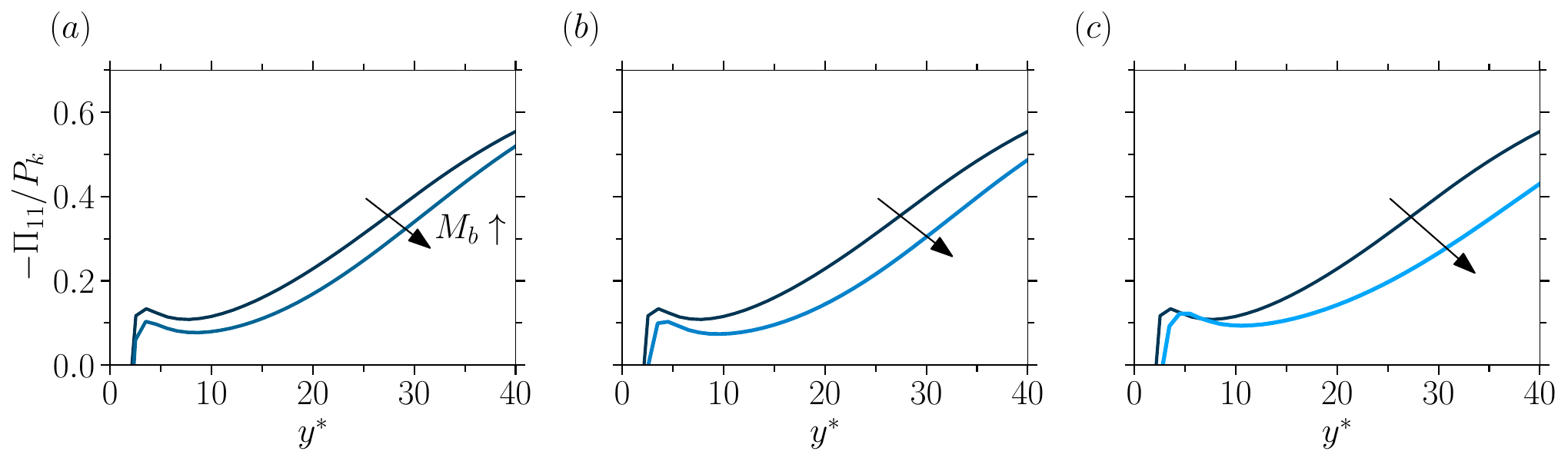}
	\caption{Wall-normal distributions of the streamwise pressure-strain correlation ($-\Pi_{11}$) scaled by the production term ($P_{11}$) for (a) Mach 2.28, (b) Mach 3 and (c) Mach 4 cases described in table~\ref{casetab}, compared to the Mach 0.3 case.}
\label{Fig:anisops}
\end{figure}

From turbulence theory, one can argue that the change in turbulence anisotropy is due to reduced inter-component energy transfer. 
Since the negative of the streamwise pressure-strain correlation ($-\Pi_{11} = -2\,\overline{p^{\prime}\partial u^{\prime\prime}/\partial x}$) is a measure of the energy transferred from the streamwise turbulent stress to the cross-stream components, we expect it to decrease with increasing Mach number for our cases. 
To verify this, figure~\ref{Fig:anisops} shows $-\Pi_{11}$ scaled by the TKE production \citep{duan2010direct, patel2015semi, cogo2023assessment}, for (a) Mach 2.28, (b) Mach 3 and (c) Mach 4 cases, compared to the Mach 0.3 case. 
The figure clearly corroborates our claims.
We further note that $\Pi_{11}$ scaled by semi-local units ($\bar \rho u_\tau^{*3}/\delta_v^*$) also reduces for the three high-Mach-number cases compared to the Mach 0.3 case (not shown). 

\subsection{Identifying direct and indirect effects of intrinsic compressibility}\label{Sec:directindirect}

So far we have observed strong intrinsic compressibility effects on various turbulence statistics. 
Are these strong effects due to a direct contribution from the dilatational motions or due to IC effects on the solenoidal motions? 
To answer this, we apply Helmholtz decomposition to the velocity field obtained from DNS to isolate the solenoidal (divergence-free) and dilatational (curl-free) parts, namely
\begin{equation}\label{helm}
    u^{\prime\prime}_i = {u_i^s}^{\prime\prime} + {u_i^d}^{\prime\prime} .
\end{equation}
Appendix~\ref{app:helm} reports details on how the decomposition is actually performed.
Following \cite{yu2019genuine}, the turbulent stresses are then split as
\begin{equation}\label{reystresssplit}
    \widetilde{ u_i^{\prime\prime} u_j^{\prime\prime}}^* = \widetilde{ {u_i^s}^{\prime\prime} {u_j^s}^{\prime\prime}}^* + \widetilde{ {u_i^d}^{\prime\prime} {u_j^s}^{\prime\prime}}^* +\widetilde{ {u_j^s}^{\prime\prime} {u_j^d}^{\prime\prime}}^* +
\widetilde{ {u_i^d}^{\prime\prime} {u_j^d}^{\prime\prime}}^*.
\end{equation}
The terms involving dilatational motions are absent in incompressible flows, and thus any contribution from them is regarded as a \emph{direct} effect.
However, the first term on the right-hand side is also present in incompressible flows. Thus, any effect of compressibility on this term will be regarded as an \emph{indirect} effect. 

\begin{figure}
	\centering	\includegraphics[width=1\textwidth]{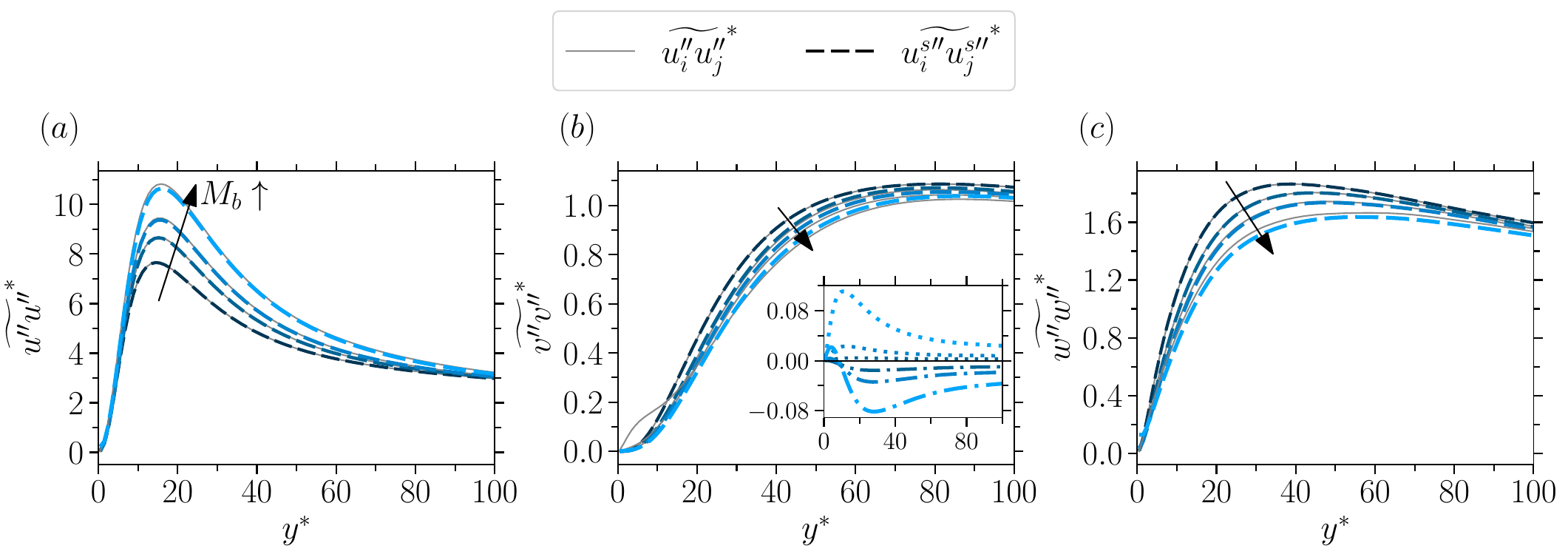}
	\caption{Wall-normal distributions of the total and solenoidal (a) streamwise, (b) wall-normal and (c) spanwise turbulent stresses as per equation~\eqref{reystresssplit}, for the cases described in table~\ref{casetab}. \textit{Inset:} profiles of the terms $\widetilde{ {v^d}^{\prime\prime} {v^d}^{\prime\prime}}^*$ (dotted) and $\widetilde{ {v^s}^{\prime\prime} {v^d}^{\prime\prime}}^*$ (dash-dotted).}
\label{Fig:normalstresssol}
\end{figure}
Figure~\ref{Fig:normalstresssol} shows the first term on the right-hand side of equation~\eqref{reystresssplit}, associated with solenoidal velocity fluctuations, for the normal turbulent stresses. They are seen to almost overlap with the total turbulent stresses, which is shown in grey. 
This implies that any change in the total stresses as a function of the Mach number is reflected in their respective solenoidal components, and thus intrinsic compressibility effects on turbulence statistics are mainly indirect. 
The collapse of the total and solenoidal stresses also implies that the correlations involving ${u_i^d}^{\prime\prime}$ are small. However, there are some exceptions, particularly the terms $\widetilde{ {v^d}^{\prime\prime} {v^d}^{\prime\prime}}^*$ and $\widetilde{ {v^s}^{\prime\prime} {v^d}^{\prime\prime}}^*$, that can have large contributions in the near-wall region as shown in the inset of figure~\ref{Fig:normalstresssol}(b). Negative values of $\widetilde{ {v^s}^{\prime\prime} {v^d}^{\prime\prime}}^*$ physically represent opposition of solenoidal motions (sweeps/ejections) from dilatational wall-normal velocity. This opposition was first observed by \cite{yu2019genuine}, and plays a key role in the forthcoming discussion.

\section{Weakening of the quasi-streamwise vortices}\label{Sec:4}

\begin{figure}
	\centering	\includegraphics[width=1\textwidth]{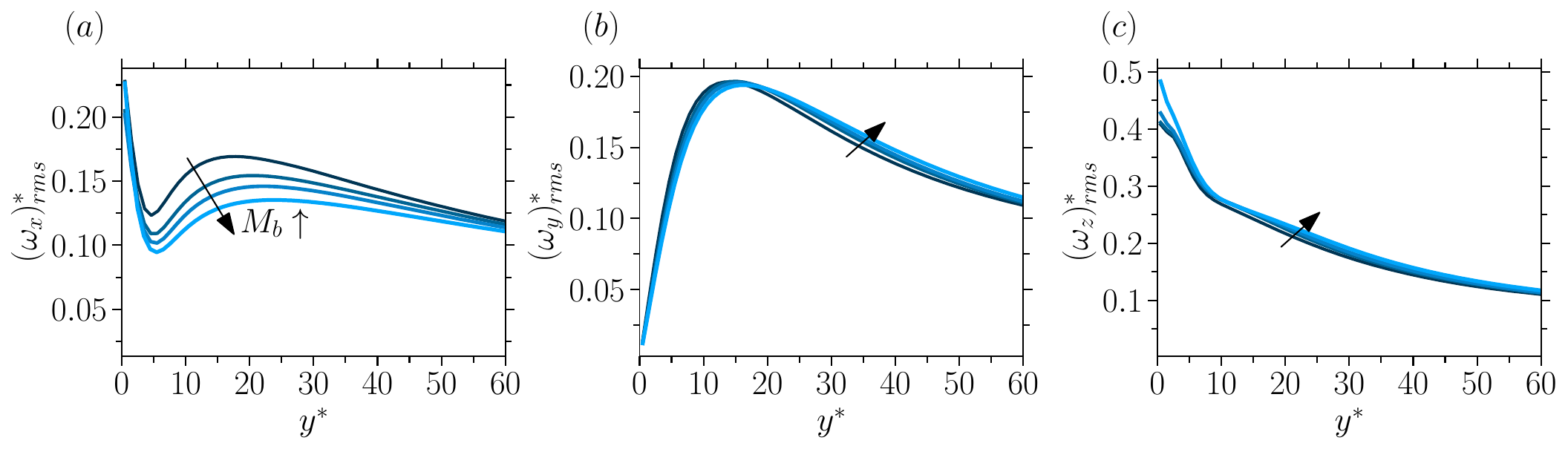}
	\caption{Wall-normal distributions of the root-mean-square of (a) streamwise, (b) wall-normal, and (c) spanwise vorticity fluctuations, scaled by $u_\tau^*/\delta_v^*$, for the cases described in table~\ref{casetab}.}
\label{Fig:vortrms}
\end{figure}

Quasi-streamwise vortices play an important role in transferring energy from the streamwise to the wall-normal and spanwise components~\citep{jeong1997coherent}. Thus, any reduction in this inter-component energy transfer (see figure~\ref{Fig:anisops}), and hence any weakening of the wall-normal and spanwise velocity fluctuations (see figure~\ref{Fig:normalstress}) is directly related to the weakening of those vortices. 
To verify this claim, the root-mean-square of the streamwise vorticity is shown in figure~\ref{Fig:vortrms}(a). This quantity indeed decreases with increasing Mach number, implying weakening of the quasi-streamwise vortices. In contrast, the root-mean-square of the wall-normal and spanwise vorticity shows a weak Mach number dependence, as seen in figure~\ref{Fig:vortrms}(b)~and~(c). 

\citet{choi1994active} showed that active opposition of sweeps and ejections is effective in weakening the quasi-streamwise vortices.
As noted in \S\ref{Sec:directindirect}, a similar opposition also occurs spontaneously in compressible flows, in which solenoidal motions like sweeps and ejections are opposed by wall-normal dilatational motions.

To explain the physical origin of near-wall opposition of sweeps and ejections, and hence the weakening of the quasi-streamwise vortices, we perform a conditional averaging procedure that identifies shear layers. Shear layers are in fact inherently associated with quasi-streamwise vortices, being formed as a consequence of sweeps and ejections initiated by those vortical structures~\citep{jeong1997coherent}.  
To educe shear layers, we rely on the variable interval space averaging (VISA) technique introduced by \cite{kim1985turbulence}, which is the spatial counterpart of the variable interval time averaging (VITA) technique developed by \cite{blackwelder1976wall}. Since only the solenoidal motions carry the imprint of incompressible turbulent structures, like shear layers, the VISA detection criterion is directly applied to the solenoidal velocity field. More details on the implementation of the VISA technique are provided in Appendix~\ref{app:visasteps}.

\subsection{Results from the variable interval space averaging technique}
\begin{figure}
	\centering	\includegraphics[height=0.85\textheight, width=0.9\textwidth]{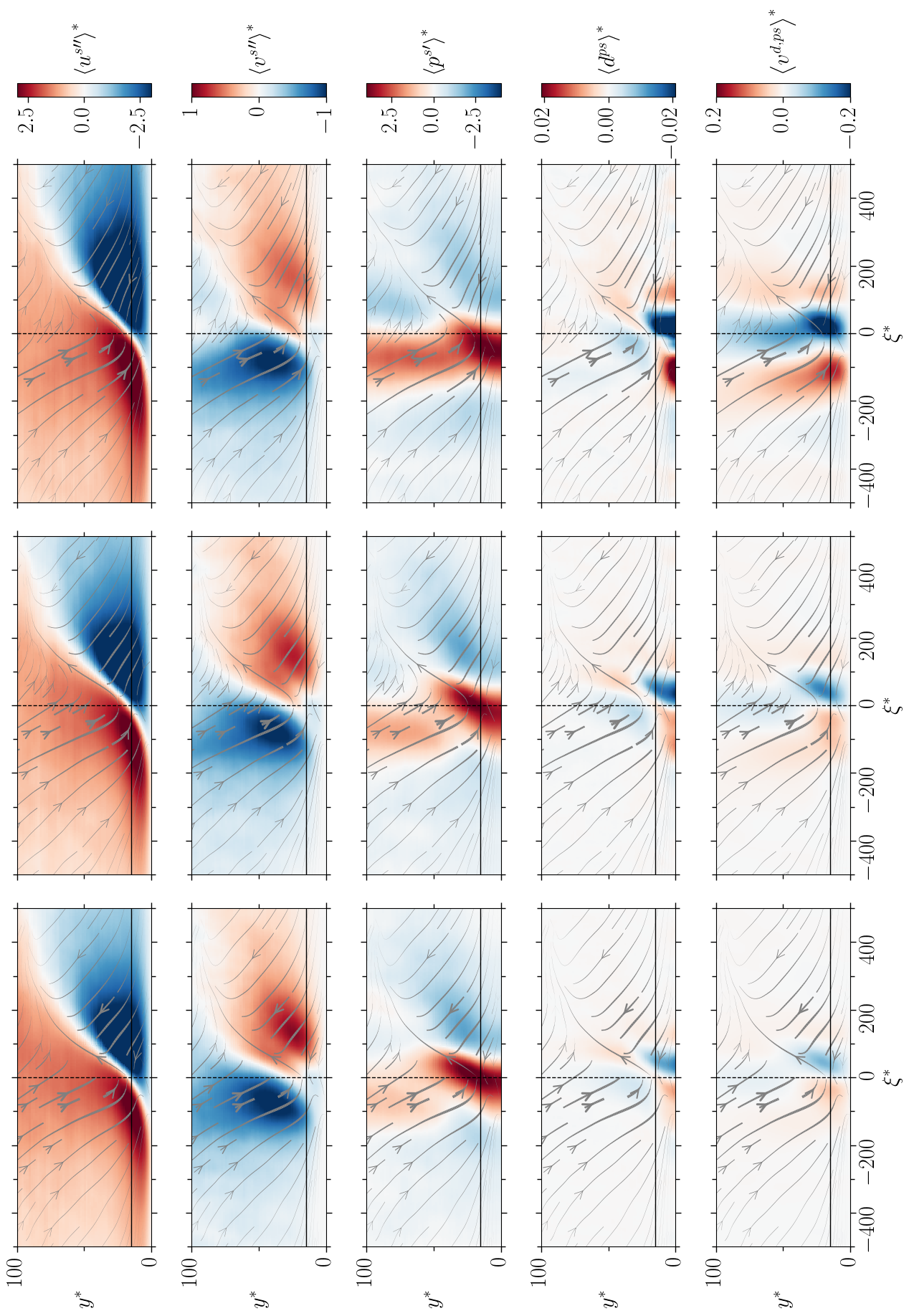}
	\caption{Conditionally averaged quantities, based on VISA applied to streamwise velocity fluctuations at $y^*\approx 15$ (see Appendix~\ref{app:visasteps}), for the Mach~2.28 (left column), Mach~3 (centre column), and Mach~4 (right column) cases in table~\ref{casetab}. The $\xi^*-y^*$ planes are taken at the centre of the shear layer 
 ($\zeta^*=0$). The velocity contours (first, second and fifth rows) are scaled by the semi-local friction velocity $u_\tau^*$, the pressure contours (third row) are scaled by $\tau_w$, and the dilatation contours (fourth row) are scaled by $u_\tau^*/\delta_v^*$.
 The overlaying streamlines are constructed using $\left<{u^s}^{\prime\prime}\right>^*$ and $\left<{v^s}^{\prime\prime}\right>^*$, and their thickness is scaled by the magnitude of $\left<{v^s}^{\prime\prime}\right>^*$. The solid black line indicates $y^*\approx 15$ and the dashed black line indicates $\xi^*=0$.}
 \label{Fig:VISA}
\end{figure}

Figure~\ref{Fig:VISA} shows the conditionally averaged $\xi^*-y^*$ planes, at $\zeta^*=0$, of various quantities for the Mach~2.28, 3 and 4 cases, only considering acceleration events. A similar plot with deceleration events is not shown since they are much less frequent~\citep{johansson1987generation}. $\xi$ and $\zeta$ indicate streamwise and spanwise coordinates, respectively, centred at the locations of the detected events.

The first row in figure~\ref{Fig:VISA} shows the contours of the conditionally averaged solenoidal streamwise velocity fluctuations $\left<{u^s}^{\prime\prime}\right>^*$, which clearly represent a shear layer. 
The second row of the plot shows the contours of the conditionally averaged solenoidal wall-normal velocity fluctuations $\left<{v^s}^{\prime\prime}\right>^*$.
Positive streamwise velocity fluctuations are associated with negative wall-normal fluctuations, resulting in a sweep event. Similarly, negative streamwise fluctuations are associated with positive wall-normal velocity, resulting in an ejection event. For greater clarity, we also show streamlines constructed using $\left<{u^s}^{\prime\prime}\right>^*$ and $\left<{v^s}^{\prime\prime}\right>^*$, with their thickness being proportional to the local magnitude of $\left<{v^s}^{\prime\prime}\right>^*$.

Similar to the velocity field, we also split pressure into solenoidal and dilatational parts, namely
\begin{equation}
    {p}^\prime = {p^s}^\prime + {p^d}^\prime.
\end{equation}
Unlike the Helmholtz decomposition for velocities, this splitting is not unique.
In this work, we adhere to the definition of solenoidal pressure given for homogeneous flows by
\cite{ristorcelli1997pseudo,jagannathan2016reynolds,wang2017spectra}, which we extend to inhomogeneous flows as follows:
\begin{equation}\label{Eq:pre_sol}
\frac{\partial^2 {p^s}^\prime}{\partial x_i\partial x_i}  = -
\frac{\partial ( \overline{\rho} {u_i^s}^{\prime\prime} {u_j^s}^{\prime\prime} - \overline{\rho} \overline{{u_i^s}^{\prime\prime} {u_j^s}^{\prime\prime}}  )}{\partial x_i \partial x_j}- 2 \overline{\rho} \frac{d\widetilde{u}}{dy} \frac{\partial {v^s}^{\prime\prime}}{\partial x}.
\end{equation}
This part of the pressure field is also referred to as pseudo-pressure~\citep{ristorcelli1997pseudo}, as it propagates with the flow speed. 
Looking at the source terms on the right-hand side of equation~\eqref{Eq:pre_sol}, the solenoidal pressure can be interpreted as being generated from vortices and shear layers, similar to incompressible flows~\citep{bradshaw1981note}. 
 
\begin{figure}
	\centering	\includegraphics[width=1
\textwidth]{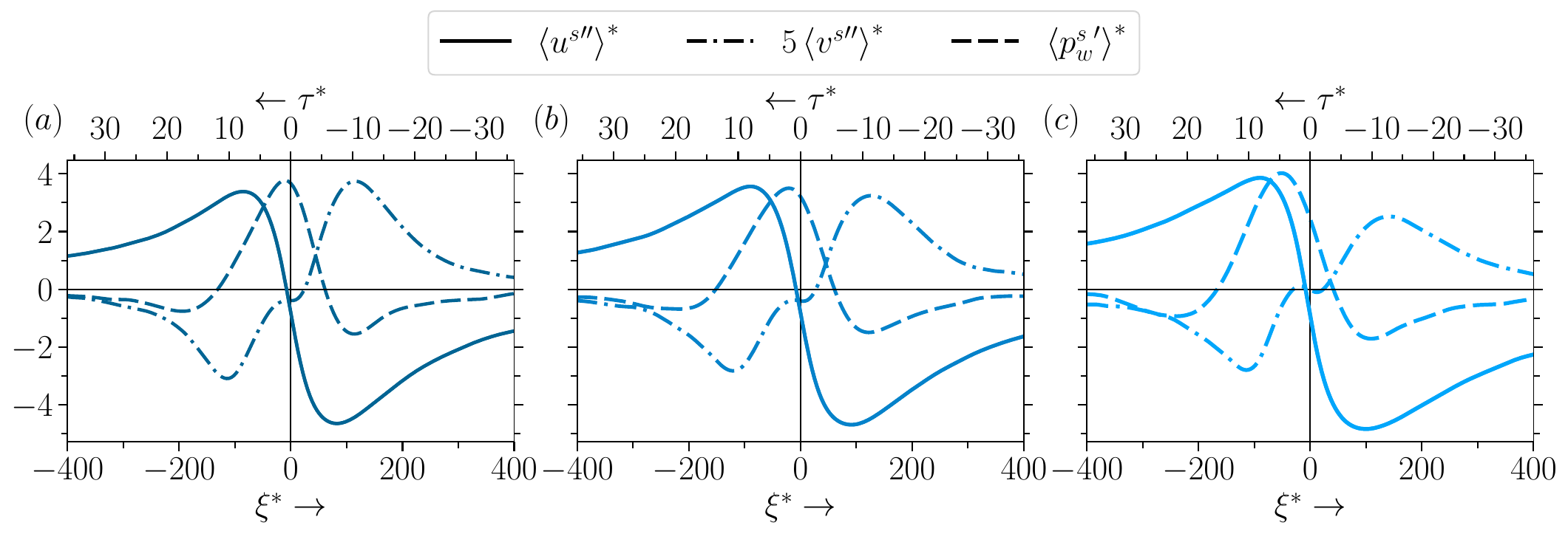}
	\caption{Conditionally averaged profiles of solenoidal streamwise and wall-normal velocities at $y^*\approx 15$, and wall pressure as a function of space ($\xi^*$, at $\zeta^*$=0; bottom-axis) and time ($\tau^* = \tau/(u_\tau^*/\delta_v^*)$; top-axis), for (a) Mach~2.28, (b) Mach~3 and (c) Mach~4 cases in table~\ref{casetab}.}
\label{Fig:temporal}
\end{figure}

The third row of figure~\ref{Fig:VISA} shows the conditionally averaged solenoidal pressure as per equation~\eqref{Eq:pre_sol}. Clearly, the pressure maxima occur approximately in between the high-velocity regions, which suggests a phase shift between velocity and pressure. To shed further light on this point, in figure~\ref{Fig:temporal} we plot the wall-normal velocity at $y^*\approx15$, and the solenoidal pressure at the wall as a function of the streamwise coordinate ($\xi^*$).
Since the wall pressure is mainly contributed by the buffer-layer eddies  \citep{kim1989structure,johansson1987generation,kim1993propagation,luhar2014structure}, its convection velocity is 
comparable with the speed of the buffer-layer coherent structures~\citep{kim1993propagation}. Using this information and Taylor's hypothesis, one can transform the spatial axis in figure~\ref{Fig:temporal} to a temporal axis ($\tau$) by taking the mean velocity at $y^*\approx 15$ as the propagation velocity. Reading figure~\ref{Fig:temporal} using the temporal axis (axis on the top), we note that the high negative sweep velocity corresponds to a high negative rate of change of the wall pressure, and likewise for the ejection velocity, i.e.,
\begin{equation}\label{Eq:vsdpdt}
    \frac{\partial \left<{p^s_w}^\prime\right>^*}{\partial \tau^*} \sim \left<{v^s}^{\prime\prime}\right>^*.
\end{equation}
Similar observations were made by~\cite{johansson1987generation}, using the VITA technique, and by~\cite{luhar2014structure}, using the resolvent analysis. 
Other interesting observations can be made from figure~\ref{Fig:temporal}. First, the magnitude of the conditionally averaged streamwise fluctuations increases, whereas the magnitude of the conditionally averaged wall-normal fluctuations decreases with increasing Mach number, as also seen in the first two rows of figure~\ref{Fig:VISA}. This is consistent with the strengthening of the streamwise and weakening of the wall-normal turbulent stresses observed in figure~\ref{Fig:normalstresssol}. Second, the wall pressure maximum shifts upstream with increasing Mach number, as also seen in the third row of figure~\ref{Fig:VISA}. While we know that such shift is attributed to the Mach number dependence of the solenoidal motions that contribute to the source terms in equation~\eqref{Eq:pre_sol}, at the moment we cannot provide a detailed explanation for this and leave it for future studies. 

After establishing the relation 
between the solenoidal wall-normal velocity and the rate of change of the solenoidal pressure in equation~\eqref{Eq:vsdpdt}, 
we continue in our attempt to relate the solenoidal and the dilatational velocity fields. 
For that purpose, we first isolate the dilatation generated from the solenoidal pressure\textemdash also referred to as `pseudo-sound' dilatation (superscript $ps$) in the literature \citep{ristorcelli1997pseudo, wang2017spectra}\textemdash, as follows
\begin{equation}
d^{ps} \approx \frac{-1}{\gamma \bar {P}} \left( \frac{\partial {p^s}^{\prime}}{\partial t} + u_j \frac{\partial {p^s}^{\prime}}{\partial x_j} \right). 
\end{equation}
Pseudo-sound dilatation represents the volume changes of fluid elements caused by pressure changes associated with solenoidal turbulent structures such as vortices and shear layers. Normalization by the wall shear stress yields 
\begin{equation}\label{dpstauwP}
d^{ps} \approx \frac{-\tau_w}{\gamma \bar P} \left( \frac{\partial {p^s}^{\prime*}}{\partial t} + u_j \frac{\partial {p^s}^{\prime*}}{\partial x_j} \right), 
\end{equation}
where the factor ${\tau_w}/({\gamma \bar P})$ is equal to the square of the semi-local friction Mach number for ideal gas flows. Using $M_\tau^*\approx M_\tau$ (see equation~\eqref{Eq:Mtau} and figure~\ref{Fig:prms}), we then rewrite equation~\eqref{dpstauwP} as
\begin{equation}\label{Eq:dps}
d^{ps} \approx -M_\tau^{2}\left( \frac{\partial {p^s}^{\prime*}}{\partial t} + u_j \frac{\partial {p^s}^{\prime*}}{\partial x_j} \right).
\end{equation}
According to the pseudo-sound theory \citep{ristorcelli1997pseudo}, the inner-scaled scaled solenoidal pressure is assumed to be unaffected by compressibility effects. Thus, from equation~\eqref{Eq:dps}, one would expect $d^{ps}$ to increase with the square of the friction Mach number. 
However, as noted in the discussion following figure~\ref{Fig:temporal}, the solenoidal motions change as a function of the Mach number, thereby affecting the solenoidal pressure as per equation~\eqref{Eq:pre_sol}. This suggests that $d^{ps}$ could increase with an exponent that is close to two but not necessarily equal to two. 
To assess the correct scaling, 
in table~\ref{scaletab} we report
the root-mean-square of $d^{ps}$ at the wall.
Data fitting yields $d^{ps} \sim M_\tau^{2.42}$, hence close to what was suggested by equation~\eqref{Eq:dps}. 

\begin{table}
\centering
\begin{tabular}{m{2.4cm} m{1.5cm} m{1.5cm} m{1.5cm} m{1.5cm} m{1.5cm} m{1.5cm}}
Case name &  $M_{b_w}$ &  $M_{\tau}$ &$\left(d_w^{ps}\right)^*_{rms}$& $\left(v_p^{d}\right)^*_{rms}$   & $\left(v_p^{d,ps}\right)^*_{rms}$ & $\left(v_p^{d,nps}\right)^*_{rms}$\\ \hline 
Mach  2.28 &2.28 & 0.1185    & 0.0096& 0.066 &0.047 & 0.059 \\ 
Mach  3   &3    & 0.1526    & 0.0160&  0.153 &0.078 &0.140\\ 
Mach  4   &4    & 0.1968    & 0.0311&  0.332 & 0.150 & 0.323\\ \hline
&&$b$ & 2.42&  3.1& 2.37& 3.3\\
\end{tabular}
\captionof{table}{Root-mean-square ($rms$) of the pseudo-sound dilatation at the wall and the peak $rms$ value of the total, pseudo-sound and non-pseudo-sound wall-normal dilatational velocities. `$b$' is the exponent obtained from power-law fitting ($a M_\tau^b$) of the data.}
\label{scaletab}
\end{table}

Continuing on our path to relate solenoidal and dilatational motions, close to the wall we can write
\begin{equation}\label{Eq:dpswall}
 {d_w^{ps}}^* \approx -M_\tau^2 \frac{\partial {p_w^s}^{\prime*}}{\partial t^*},   
\end{equation}
where ${d_w^{ps}}^* = {d_w^{ps}}/(u_\tau^*/\delta_v^*)$. This equation, when conditionally averaged and combined with equation~\eqref{Eq:vsdpdt}, leads to 
\begin{equation}\label{Eq:vsdps}
   \left<d_w^{ps}\right>^* \sim - M_\tau^2 \left<{v^s}^{\prime\prime}\right>^*.   
\end{equation}
Using this result, we expect positive dilatation events (expansions) to be mainly associated with sweeps and negative dilatation events (compressions) to be associated with ejections.
The fourth row in figure~\ref{Fig:VISA} shows the contours of conditionally averaged pseudo-sound dilatation defined in equation~\eqref{Eq:dps}. Consistent with our expectation, positive dilatation is indeed found to be associated with sweeps and negative dilatation with ejections, and its magnitude increases with the Mach number. 

Having related the pseudo-sound dilatation and the solenoidal velocity in equation~\eqref{Eq:vsdps}, the next step is to introduce the pseudo-sound dilatational velocity as
\begin{equation}\label{Eq:phips}
    \begin{aligned}
\frac{\partial^2 \phi^{ps} }{\partial x_j \partial x_j} &= d^{ps},\\
v^{d,ps} &= \dfrac{\partial \phi^{ps}}{\partial y},
            \end{aligned}
\end{equation}
where $\phi^{ps}$ is the scalar potential. Note that this equation is similar to equations~\eqref{udphi}~and~\eqref{phihelm} used to solve for the total dilatational velocity, as reported in Appendix~\ref{app:helm}.
Based on equation~\eqref{Eq:phips}, one would expect $v^{d,ps}$ to increase with the Mach number at a similar rate as $d^{ps}$. 
Power-law fitting of the data reported in table~\ref{scaletab} indeed yields $v^{d,ps} \sim M_\tau^{2.37}$, hence close to what was found for $d^{ps}$.

Equation~\eqref{Eq:phips} stipulates that the conditionally averaged pseudo-sound dilatational velocity in the buffer layer should be proportional to and in phase with the dilatation at the wall. 
Thus, we can write
\begin{equation}\label{vdpsdps}
   \left<{v^{d,ps}}\right>^* \sim \left<{d_w^{ps}}\right>^*.  
   \end{equation}
Using equation~\eqref{vdpsdps} and ~\eqref{Eq:vsdps} we can finally develop a relation between the solenoidal and the pseudo-sound dilatational velocity, namely 
\begin{equation}\label{Eq:vsvdps}
   \left<{v^{d,ps}}\right>^* \sim - M_\tau^2 \left<{v^s}^{\prime\prime}\right>^*.   
\end{equation}
In our opinion, this relation is quite meaningful as it theoretically supports near-wall opposition of sweeps and ejections by dilatational motions. Moreover, it suggests that the opposition effect should approximately increase with the square of $M_\tau$. 

\begin{figure}
	\centering	\includegraphics[width=1
\textwidth]{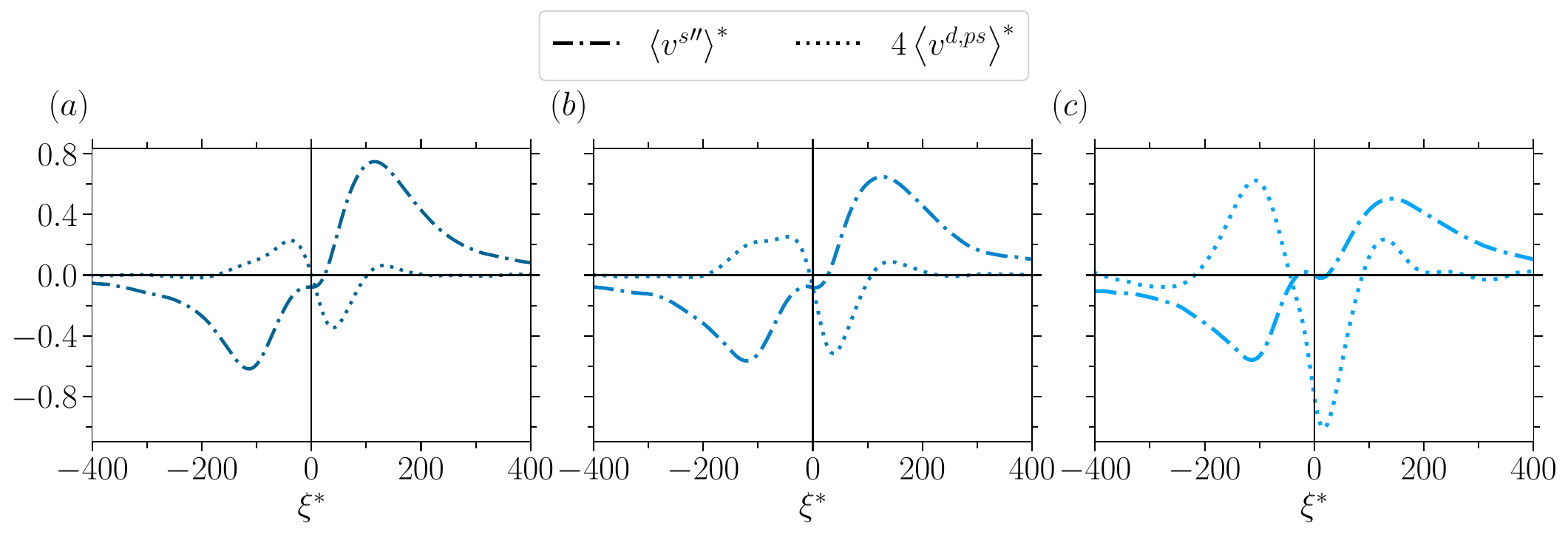}
	\caption{Conditionally averaged profiles of solenoidal and pseudo-sound dilatational 
    wall-normal velocities at $y^*\approx 15$ as a function of $\xi^*$ (at $\zeta^*$=0) for (a) Mach 2.28, (b) Mach 3 and (c) Mach 4 cases in table~\ref{casetab}.}
\label{Fig:vsvdline}
\end{figure}

In order to verify this, the final row in figure~\ref{Fig:VISA} reports the conditionally averaged contours of the pseudo-sound wall-normal dilatational velocity given in equation~\eqref{Eq:phips}. As suggested from equation~\eqref{vdpsdps}, the contours of $v^{d,ps}$ appear to be in phase with those of $d^{ps}$.
Thus, consistent with the observations made for the pseudo-sound dilatation, the wall-normal dilatational velocity is positive during sweeps and negative during ejections, and its magnitude increases with the Mach number. This opposition is also clearly seen in figure~\ref{Fig:vsvdline}, which shows the conditionally averaged profiles of ${v^s}^{\prime\prime}$ and $v^{d,ps}$ at $y^*\approx 15$. Additionally, in figures~\ref{Fig:VISA}~and~\ref{Fig:vsvdline} we note that the pseudo-sound dilatational velocity contour (or profile) shifts upstream (leftward) with increasing Mach number. This is due to the upstream shift in the pressure contour mentioned above. 

\begin{figure}
	\centering	\includegraphics[width=1\textwidth]{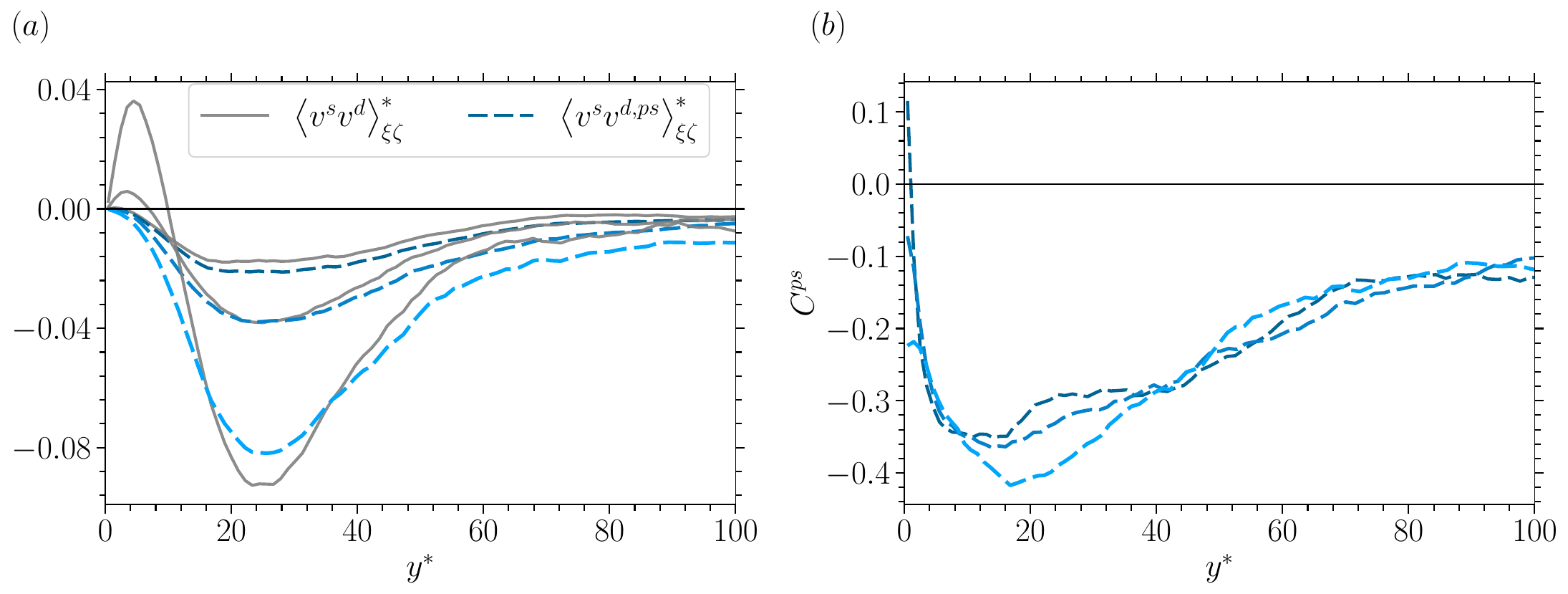}
	\caption{(a) Conditionally averaged and integrated [equation~\eqref{xzavg}] correlations between solenoidal and dilatational velocities. (b) Conditionally averaged pseudo-sound correlation coefficient ($C^{ps}$) as defined in equation~\eqref{corrcoeff}.}
\label{Fig:vsvd}
\end{figure}
To further quantify the opposition effect, we analyse the conditionally averaged correlation between solenoidal and pseudo-sound dilatational wall-normal velocity, i.e. $\left<v^sv^{d,ps}\right>$. The correlation is integrated over a window of 300 viscous units in the streamwise direction and 40 viscous units in the spanwise direction \citep{johansson1991evolution}, at each wall-normal location as   \begin{equation}\label{xzavg}
\left<{v^s}^{\prime\prime}v^{d,ps}\right>_{\xi\zeta}(y^*)= \int_{\zeta^*=-20}^{20}\int_{\xi^*=-150}^{150} \left<{v^s}^{\prime\prime}v^{d,ps}\right>(\xi^*,y^*,\zeta^*) d\xi^* d\zeta^*.
\end{equation}
The integrated correlation, scaled by the squared semi-local friction velocity, is reported in figure~\ref{Fig:vsvd} with dashed lines. Figure~\ref{Fig:vsvd} also shows the pseudo-sound correlation coefficient defined as
\begin{equation}\label{corrcoeff}
C^{ps}=\frac{\left<{v^s}^{\prime\prime}v^{d,ps}\right>_{\xi \zeta}}{\sqrt{\left<{v^s}^{\prime\prime} {v^s}^{\prime\prime}\right>_{\xi \zeta}\left<v^{d,ps} v^{d,ps}\right>_{\xi \zeta}}}.
\end{equation}
The correlation and its coefficient are negative as expected. The magnitude of the correlation increases approximately with the square of Mach number, as expected. 
However, the correlation coefficient almost collapses across all Mach numbers. 

\begin{figure}
	\centering	\includegraphics[width=1\textwidth]{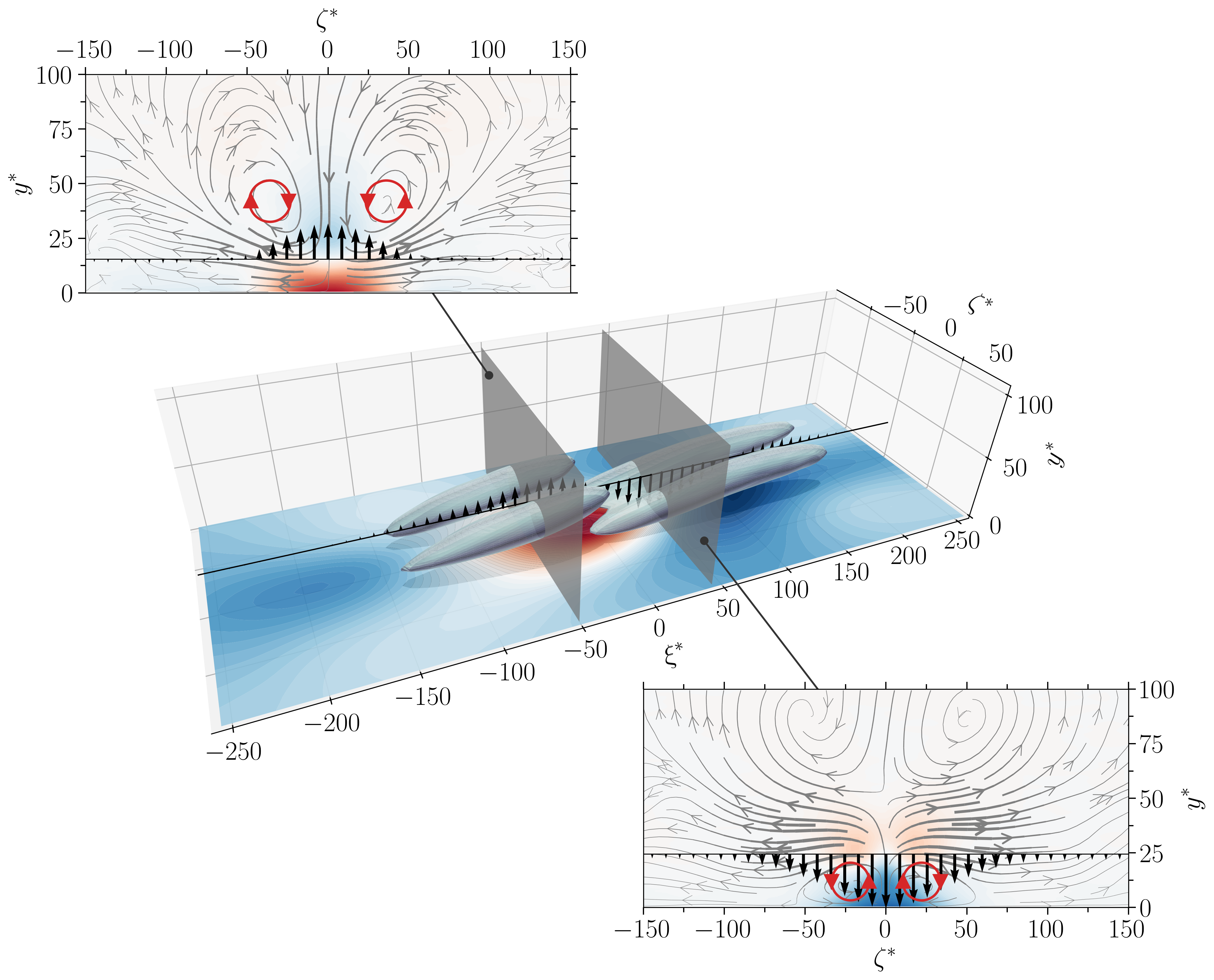}
	\caption{Opposition of sweeps and ejections by wall-normal pseudo-sound dilatational velocity in the context of quasi-streamwise vortices. 
 The shaded three-dimensional isosurfaces represent quasi-streamwise vortices identified by applying the Q-criterion to the conditionally averaged velocity field. Their shadows are also plotted on the wall below, showing that the vortices are inclined and tilted. Underneath the vortices, the contours of solenoidal wall pressure are shown. 
 The transparent planes mark regions of high rate of change of wall pressure and hence high wall-normal pseudo-sound dilatational velocity $\left<v^{d,ps}\right>^*$ (see discussion related to equations~\eqref{Eq:vsdpdt}~-~\eqref{Eq:vsvdps}). The arrows between the vortices indicate $\left<v^{d,ps}\right>^*$ as a function of $\xi^*$ at $\zeta^*=0$ and $y^*\approx20$. Note that the line along which the arrows are plotted is slightly shifted away from the wall for better visibility. \textit{Insets:} contours of pseudo-sound dilatation $\left<d^{ps}\right>^*$ along the transparent planes, overlaid with the streamlines generated by quasi-streamwise vortices. These streamlines are constructed using the wall-normal and spanwise solenoidal velocities, i.e. $\left< v^{s}\right>^*$ and $\left< w^{s}\right>^*$, with their thickness being proportional to the magnitude of the local planar velocity.
 $\left<v^{d,ps}\right>^*$ at $y^*\approx15$ and $y^*\approx25$ is also shown using arrows in the left and right planes, respectively. These wall-normal locations correspond to the maximum value of $\left<v^{d,ps}\right>^*$ in those planes. 
    The red and blue colours in the contour plots indicate positive and negative values, respectively.
    An interactive version of this figure can be accessed \href{https://fluid-dynamics-of-energy-systems-team.github.io/IntrinsicCompressPlot/}{here}.}
\label{Fig:schematic}
\end{figure}

The association of the opposition effect with the quasi-streamwise vortices is visualised in figure~\ref{Fig:schematic} for the Mach~2.28 case, all other cases being qualitatively similar.
Indeed, the figure insets illustrate that sweeps and ejections initiated by quasi-streamwise vortices are opposed by the near-wall pseudo-sound dilatational velocity, thereby resulting in their weakening.

\subsection{Role of non-pseudo-sound dilatational velocity in near-wall opposition}

So far we have looked into the pseudo-sound dilatational velocity and provided an explanation for why they are out-of-phase with respect to the solenoidal motions. However, from table~\ref{scaletab}, we see that the peak root-mean-square value of $v^{d,ps}$ is much smaller than that of the total dilatational velocity. 
Hence, a large portion of the dilatational velocity and its correlation (if any) with the solenoidal velocity is still unexplained. To address this point, figure~\ref{Fig:vsvd}(a) shows the integrated correlation between solenoidal and total dilatational velocities, i.e. $\left<{v^s}^{\prime\prime}v^{d}\right>^*_{\xi \zeta}$, denoted by solid grey lines.  
Except very close to the wall, the total and pseudo-sound correlations almost overlap. This implies that the contribution from the remaining portion of the dilatational velocity, referred to as the `non-pseudo-sound' component and given by
\begin{equation}
v^{d,nps} = v^{d} - v^{d,ps} ,
\end{equation} 
is small. In other words, 
despite being stronger in magnitude than the pseudo-sound component, 
the non-pseudo-sound dilatational velocity does not play an important role in opposing sweeps and ejections.

\begin{figure}
	\centering	\includegraphics[width=1\textwidth]{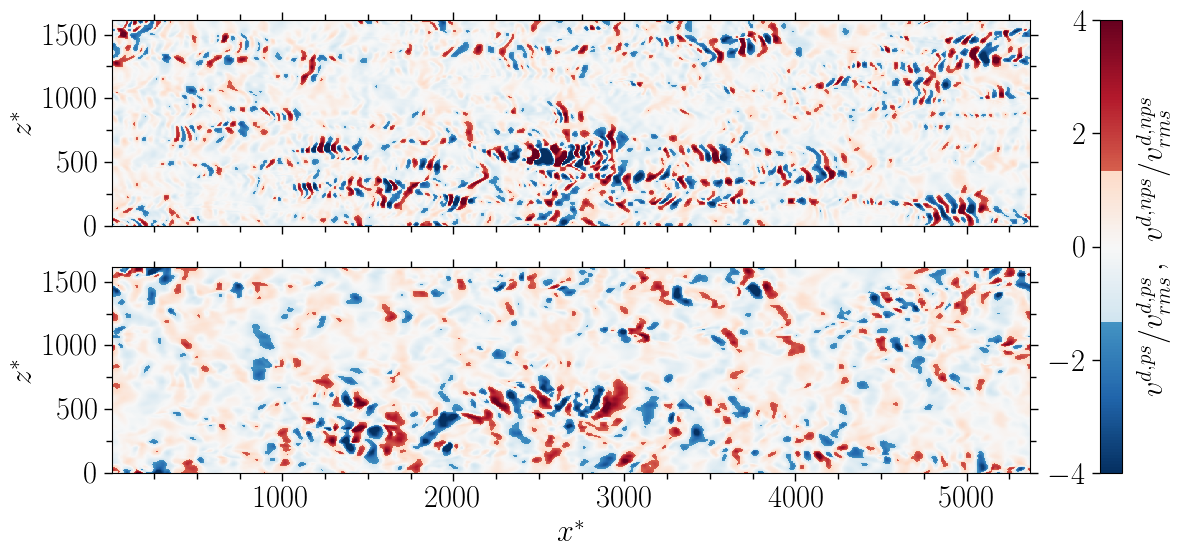}
	\caption{Instantaneous $x^*-z^*$ planes at $y^*\approx 11$ of (top) the non-pseudo-sound and (bottom) the pseudo-sound wall-normal dilatational velocities (see the text for definitions) scaled by their respective root-mean-squares for the Mach 3 case in table~\ref{casetab}. Note that, for clarity, the colour bar is adjusted such that structures stronger than 1.33 times the root-mean-square value are highlighted.}
\label{Fig:psnps}
\end{figure}

Before concluding, we would like to comment on the travelling wave-packet-like structures, first identified by \cite{yu2019genuine} and later studied in \cite{yu2020compressibility,yu2021compressibility,tang2020near,gerolymos2023scaling,yu2024generation}.
Figure~\ref{Fig:psnps} shows the $x^*-z^*$ plane with the instantaneous contours of the pseudo-sound and non-pseudo-sound dilatational velocity at $y^*\approx 11$, for the Mach 3 case in table~\ref{casetab}. The wave-packet structures are predominantly present in the non-pseudo-sound component, whereas the pseudo-sound component shows a spotty structure similar to that observed for the streamwise gradient of wall pressure in incompressible flows~\citep{kim1989structure}. Combining the observation above that the non-pseudo-sound component hardly contributes to the opposition effect, and that the wave-packet-like structures are present mainly in this component, one can argue that these structures do not play an important role in opposing sweeps and ejections. 

\section{Conclusions}
\label{Sec:5}
\begin{figure}
	\centering	\includegraphics[width=1\textwidth]{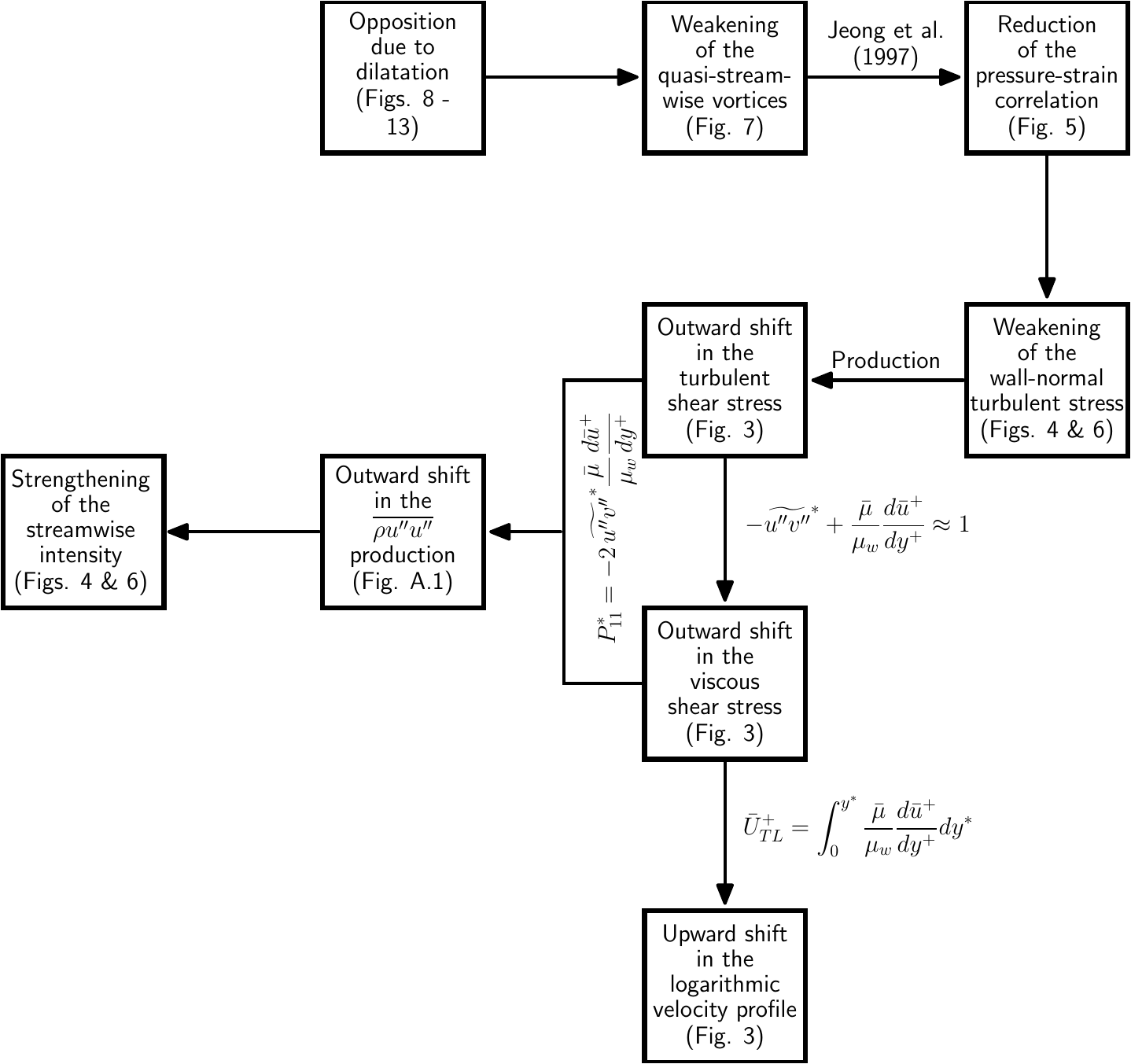}
	\caption{A graphical summary of the present findings. Note that the arrows are meant to indicate the chain of arguments made in this paper, not relations of causality.}
\label{Fig:summary}
\end{figure}

In this paper, we have attempted to provide an explanation for the underlying mechanism through which intrinsic compressibility effects modulate the near-wall dynamics of turbulence. To rigorously assess these effects, we have devised four DNS cases of fully developed high-Mach-number channel flows with approximately constant mean properties, whereby intrinsic compressibility effects are isolated. Our findings, sketched as a flow chart in figure~\ref{Fig:summary}, are summarised as follows.

First, we have decomposed the velocity field into solenoidal and dilatational parts and educed shear layers by applying conditional averaging to the solenoidal component. 
We have noticed that there exists a streamwise phase shift between the buffer-layer sweeps and ejections that form shear layers, and the associated `solenoidal' wall pressure.
Equivalent observations were made for incompressible flows by \cite{johansson1987generation} and \cite{luhar2014structure}. By using Taylor's hypothesis, this streamwise shift in phase can be interpreted as a phase shift in time, such that regions of high positive rate of change of wall pressure correspond to regions of high positive wall-normal velocity. Similarly, regions of high negative rate of change of wall pressure correspond to the regions of high negative wall-normal velocity. 
Close to the wall, the high rate of change of the solenoidal pressure results in large dilatation values with an opposite sign (also referred to as pseudo-sound dilatation), which upon integration results in a wall-normal dilatational velocity that inherently opposes sweeps and ejections.
Since sweeps and ejections are initiated by quasi-streamwise vortices, their opposition directly affects the evolution of those vortices, causing their weakening. 
This is schematically depicted in figure~\ref{Fig:schematic}.

Interestingly, we also found that the remaining portion of the dilatational velocity (also referred to as the non-pseudo-sound component) does not play an important role in the opposition mechanism.
Moreover, we have observed that the majority of the travelling wave-packet-like structures, recently discovered in the literature, are present in this non-pseudo-sound component.

The weakening of quasi-streamwise vortices directly hinders the energy transfer from the streamwise velocity component to the other two components, resulting in an outward shift (reduction) in the wall-normal turbulent stress with increasing Mach number. Since the wall-normal motions actively contribute to the transport of momentum across mean shear, thereby generating turbulent shear stress, the outward shift in the wall-normal turbulent stress results in a corresponding outward shift in the turbulent shear stress. This reduction in the turbulent shear stress is in turn responsible for an upward shift in the logarithmic mean velocity profile \citep{hasan2023incorporating}.

A longstanding question in the compressible flow community is why the inner-scaled streamwise turbulent stress is higher in compressible flows than in incompressible flows, with similar Reynolds numbers. In this respect, our results suggest that intrinsic compressibility effects play a dominant role. Specifically, the increase in the peak value is a consequence of the outward shift in the turbulent and viscous shear stresses, since their product yields the production of the streamwise turbulence stress. This implies that the near-wall opposition mechanism outlined above is also responsible for the strengthening of the streamwise turbulence intensity. 

Some questions related to the findings made in this paper remain unanswered as of yet. First, why do the solenoidal pressure maxima shift upstream with increasing Mach number (see figures~\ref{Fig:VISA}~and~\ref{Fig:temporal})? Second, what is the Mach number scaling of the turbulence statistics presented in the paper? This could help explain the quasi-linear increase in the log-law constant observed by \citet{hasan2023incorporating}. Moreover, knowing the Mach number scaling of the peak streamwise turbulence intensity would help in developing empirical scaling laws. Third, why is the dissipation of turbulence kinetic energy, and thus the small scales of turbulence, not affected by intrinsic compressibility effects (see Appendix~\ref{Sec:streamwise})? A spectral analysis of the velocity field could shed more light on this important issue.

\appendix
\renewcommand\thefigure{\thesection.\arabic{figure}}
\section{Increase in the streamwise turbulence intensity}\label{Sec:streamwise}
\setcounter{figure}{0}

In order to explain the increase in the streamwise turbulent stress and hence in the turbulence kinetic energy, we consider the streamwise turbulent stress budget for a fully-developed compressible channel flow:
\begin{equation}
    P_{11} + \epsilon_{11} + T_{11}^{\nu} + T_{11}^{u} + \Pi_{11} + C_{11} = 0 \mathrm{,} 
    \label{eq:budget11}
\end{equation}
where
\begin{gather}\label{eq:budget11b}
P_{11}  =-2 \overline{\rho {u}^{\prime \prime} {v}^{\prime \prime}} \frac{\partial \widetilde{u}}{\partial y}, ~
\epsilon_{11}  = - 2 \overline{\tau^\prime_{1 j} \frac{\partial {u}^{\prime \prime}}{\partial x_j}},\nonumber\\ 
T^{\nu}_{11} =2 \frac{\partial}{\partial y}\left(\overline{{\tau^\prime_{1 2}} {u}^{\prime \prime}}\right), ~
T_{11}^{u} =-\frac{\partial}{\partial y}\left(\overline{\rho {u}^{\prime \prime} {u}^{\prime \prime} {v}^{\prime \prime}}\right), ~ \nonumber\\ 
\Pi_{11}  = 2 \overline{p^{\prime}\frac{\partial u^{\prime}}{\partial x}},~
C_{11}  = 2 \overline{u^{\prime \prime}}\frac{\partial \bar{\tau}_{1 2}}{\partial y}.
\end{gather}

\begin{figure}
	\centering	
 \includegraphics[width=1
\textwidth]{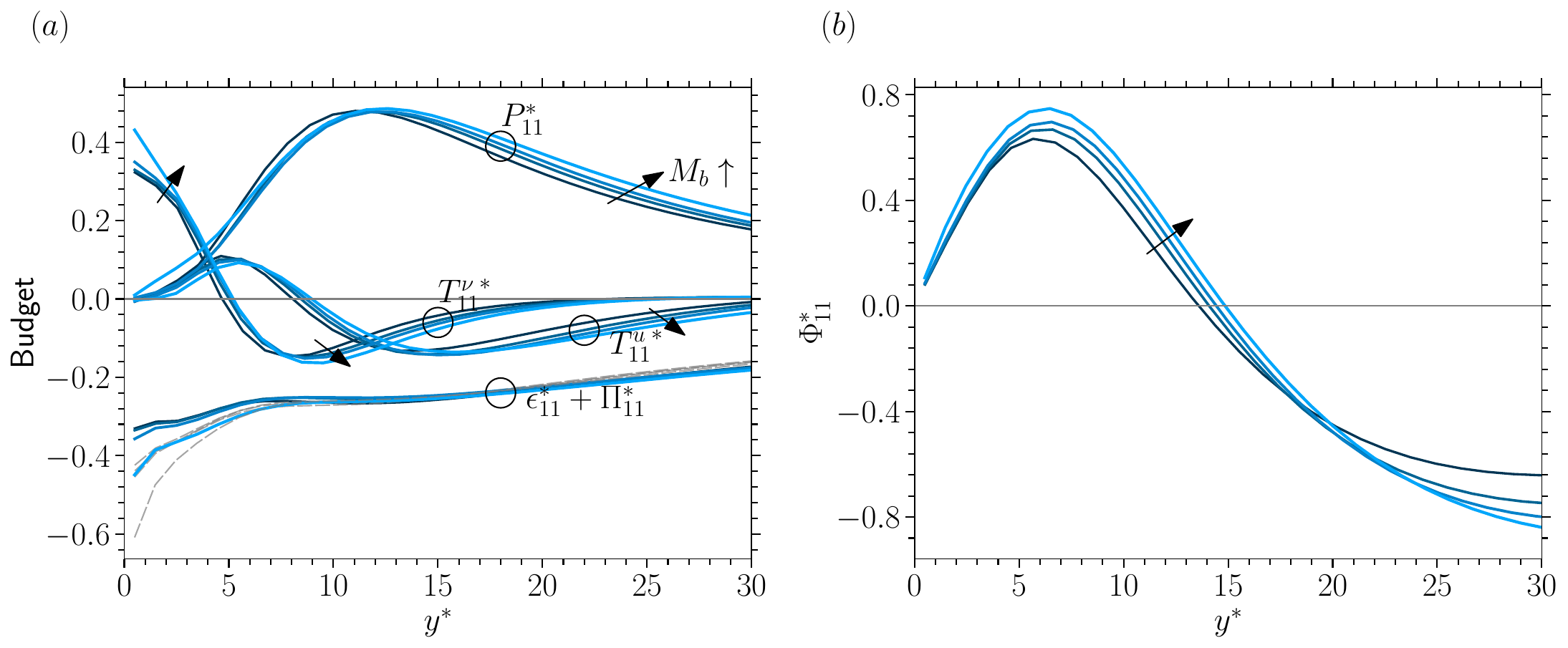}   
	\caption{Wall-normal distributions of (a) the streamwise turbulent stress budget [see equation~\eqref{eq:budget11}] scaled  in semi-local units, and (b) the sum of viscous and turbulent fluxes obtained upon integrating the semi-locally scaled viscous and turbulent diffusion terms [see equation~\eqref{eq:flux}], for the cases described in table~\ref{casetab}.} \label{Fig:bud}
\end{figure}

The distributions of the production, viscous and turbulent diffusion terms, and the sum of dissipation and pressure-strain correlation,  
are shown in figure~\ref{Fig:bud}, scaled by $\bar \rho {u_\tau^*}^3/\bar \mu$.  
The compressibility term $C_{11}$ is omitted because of its negligible magnitude.  

Three observations can be made. 
First, there is an outward shift in $P_{11}^*$ with increasing Mach number. 
Since the production term in scaled form is simply the product of the turbulent and viscous shear stresses, its outward shift is explained by the corresponding shift in the shear stresses in figure~\ref{Fig:shearstress}(b) as follows. 
Assuming that the total stress is approximately equal to $\tau_w$, such that the sum of the scaled stresses is unity, one can substitute the viscous shear by the turbulent shear stress in $P_{11}$ to obtain~\citep{pope2001turbulent}
\begin{equation}
P_{11}^* \approx -2\widetilde{ u^{\prime\prime} v^{\prime\prime}}^*\left(1+\widetilde{ u^{\prime\prime} v^{\prime\prime}}^*\right) = 2\left(-\widetilde{ u^{\prime\prime} v^{\prime\prime}}^*\right)-2\left(-\widetilde{ u^{\prime\prime} v^{\prime\prime}}^*\right)^2.
\end{equation}
Taking the derivative of $P^*_{11}$ with respect to the turbulent shear stress yields
\begin{equation}\label{Eq:dpdr}
\frac{d P_{11}^*}{d \left(-\widetilde{ u^{\prime\prime} v^{\prime\prime}}^*\right)} \approx 2- \, 4 \left(-\widetilde{ u^{\prime\prime} v^{\prime\prime}}^*\right).
\end{equation}
Between the wall and the location where $-\widetilde{ u^{\prime\prime} v^{\prime\prime}}^*$ is equal to 0.5 the derivative is positive, while it is negative above this location. 
On the other hand, from figure~\ref{Fig:shearstress}(b), we observe that the rate of change of the turbulent shear stress with the Mach number,
i.e. $\partial (-\widetilde{ u^{\prime\prime} v^{\prime\prime}}^*)/\partial M_b$, at a fixed $y^*$ is negative.
Combining these two observations, we can conclude that the rate of change of production of the streamwise turbulent stress with the Mach number, i.e. $\partial P_{11}^*/\partial M_b$, is negative close to the wall and becomes positive away from it, resulting in an effective outward shift.

Second, except very close to the wall, the sum of the two sink terms in the budget of the streamwise turbulent stress \eqref{eq:budget11}, namely $\epsilon_{11}^*$ and $\Pi_{11}^*$, show a weak Mach number dependence. 
Interestingly, the TKE dissipation ($2 \epsilon_k^*=\epsilon^*_{11}+\epsilon^*_{22}+\epsilon^*_{33}$), reported with grey dashed lines in figure~\ref{Fig:bud}, also shows marginal  dependence on the Mach number. This is consistent with the observation made by \cite{hasan2023incorporating} regarding the universality of the local Kolmogorov length scale. The universality of $\epsilon_{11}^*+\Pi_{11}^*$ and $\epsilon_{k}^*$ are related as follows.
Any Mach-number-dependent reduction in $\Pi_{11}^*$ would imply that less energy is being received by the lateral turbulent stresses, and hence, less TKE is being dissipated through the terms $\epsilon^*_{22}+\epsilon^*_{33}$. This suggests that the Mach-number-dependence of $\Pi_{11}^*$ and $\epsilon^*_{22}+\epsilon^*_{33}$ is linked, such that the universality of $\epsilon^*_{11}+\Pi^*_{11}$ is connected with the universality of the TKE dissipation.

Third, above $y^*\approx 12$, the production term is higher at higher Mach numbers, which combined with the observation that the total sink $\epsilon^*_{11}+\Pi^*_{11}$ is universal, implies more negative values of the diffusion term. This means that the surplus production is transported away from the buffer layer towards the wall. For further insight, figure~\ref{Fig:bud}(b) shows the sum of the viscous and turbulent fluxes obtained by integrating the transport terms as
\begin{equation}\label{eq:flux}
\Phi^*_{11} =   \int_0^{y^*} \left(T_{11}^{\nu*}+T_{11}^{u*}\right) d y^*, 
\end{equation}
such that positive values signify that energy is transported towards the wall, and negative values signify the opposite. As one can observe, the flux is positive close to the wall and increases with the Mach number. This implies that more energy is being carried towards the wall at higher Mach numbers. Between the wall and the peak location of the streamwise turbulence intensity, the total flux is mainly controlled by the viscous flux, which can be approximated as $d\widetilde{u^{\prime\prime}u^{\prime\prime}}^*/dy^*$. Thus, a higher positive flux at increasing Mach numbers implies a higher gradient of the streamwise turbulent stress, which results in a higher peak value upon integration.

The strengthening of the streamwise velocity fluctuations can also be explained based on a phenomenological mixing-length model. The semi-locally scaled streamwise stress can be written as 
\begin{equation}\label{uumodel}
    \left(\overline{u'u'}^*\right)^{1/2} \sim \ell^* \frac{d \bar U^*}{dy^*},
\end{equation}
where $\ell^*$ is the mixing length scaled by $\delta_v^*$, and ${d \bar U^*}/{dy^*}$ is the semi-locally scaled mean velocity gradient, which is equivalent to $d \bar U^+_{TL}/{dy^*}$. 
Note that the streamwise stress is written in the Reynolds averaged form, since we observe that the peak of both Reynolds and Favre averaged stresses increases alike (not shown), and therefore the error incurred by excluding density fluctuations from equation~\eqref{uumodel} is small. 
The mixing length is determined  as $\ell^* \sim \sqrt{\overline{v' v'}^*} \, \mathcal{T}$ \citep{durbin1991near}, where $\mathcal{T}\sim k^*/\epsilon^*$. For the present cases this definition of mixing length yields universal distributions across the Mach number range (not shown). This is because the velocity with which a fluid parcel travels reduces with increasing Mach number. However, the time scale over which it retains its streamwise momentum increases with the Mach number (due to higher TKE and almost universal dissipation), thus effectively travelling the same distance. 
Due to the universality of the mixing length, equation~\eqref{uumodel} implies that the increase in mean shear observed in figure~\ref{Fig:shearstress} is directly responsible for an increase in the peak streamwise turbulence intensity. Interestingly, an increase in the mean shear was also found to be responsible for higher production in the buffer layer (see figure~\ref{Fig:bud}) that formed the basis of our explanation above, making the phenomenological model consistent.

\section{Helmholtz decomposition of the velocity field}\label{app:helm}
\setcounter{figure}{0}

The Helmholtz decomposition in compressible flows is the representation of the velocity field as the sum of divergence-free `solenoidal' and curl-free `dilatational' components. 
This is mathematically written as
\begin{equation}\label{helmapp}
    {u}_i = {u_i^s} + {u_i^d},
\end{equation}
where superscripts `$s$' and `$d$' stand for solenoidal and dilatational components. This equation is similar to equation~\eqref{helm} in the main text, the only difference being that there the decomposition was written explicitly for the fluctuating velocity field.  

The dilatational component is computed as the gradient of a scalar potential $\phi$, namely 
\begin{equation}\label{udphi}
    {u_i^d} = \frac{\partial \phi}{\partial x_i},
\end{equation}
where $\phi$ is obtained by solving a Poisson equation as
\begin{gather}\label{phihelm}
\frac{\partial^2 \phi}{\partial x_j \partial x_j}= \frac{\partial u_i}{\partial x_i}.
\end{gather}
Equation~\eqref{phihelm} is solved using a second-order accurate FFT-based Poisson solver (see \cite{costa2018fft} for example) with
periodic boundary conditions in the streamwise and spanwise directions, and no-penetration boundary condition $\partial \phi/\partial y = 0$ (or $v^d=0$) at the wall. Note that with these boundary conditions, no-slip is not satisfied at the wall, that is $u^d$ and $w^d$ are not equal to zero. While seemingly counter-intuitive at first glance, this is not unphysical, as pointed out in \cite{sharma2023effect}.

Likewise, the solenoidal component can be obtained using the vorticity field as described in \cite{yu2019genuine} and \cite{sharma2023effect}. However, here we will make use of the fact that the total velocity field is available from the direct numerical simulation. Thus, the solenoidal field is simply computed using equation~\eqref{helmapp} as
\begin{equation}
    {u_i^s} = {u_i} - {u_i^d}.
\end{equation}

\section{Steps to perform variable interval space averaging}\label{app:visasteps}
\setcounter{figure}{0}

In this conditional average technique, strong sweep and ejection events resulting in a shear layer are said to occur when the short-space variance, given by 
\begin{equation}\label{Eq:shortspacevar}
\mathrm{var}(x,z,t) = \frac{1}{L} \int_{-\frac{L}{2}}^{\frac{L}{2} } [{u^s}^{\prime\prime}(x+s,y_{ref},z,t)]^2 \, \mathrm{d} s-\left(\frac{1}{L} \int_{-\frac{L}{2}}^{\frac{L}{2} } {u^s}^{\prime\prime}(x+s,y_{ref},z,t) \, \mathrm{d}s \right)^2,
\end{equation}
exceeds $K [u^s_{rms}(y_{ref})]^2$, where $K$ is the threshold level. Here, $y_{ref}$ is the location of the reference $x-z$ plane where the detection criteria is applied, and $L$ is the size of the averaging window, representative of the length scale of the shear layer identified by this technique \citep{johansson1987generation}. Following \citet{johansson1991evolution}, we take $K=1$, $y^*_{ref}\approx15$, and $L^*\approx200$.

\begin{figure}
	\centering	\includegraphics[width=1\textwidth]{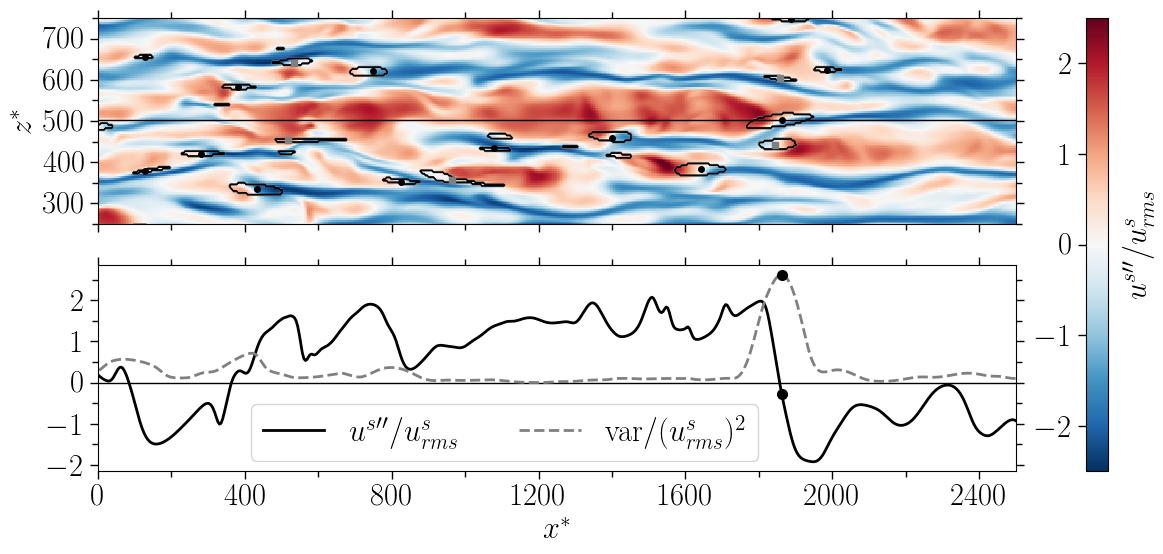}
	\caption{(Top) $x^*-z^*$ contour plot of the instantaneous solenoidal streamwise velocity fluctuations at $y^*_{ref}\approx15$ for the Mach 2.28 case. Boundaries of patches where the short-space variance exceeds the Reynolds averaged value [see equation~\eqref{Cond}] are overlaid on the contour plot. Additionally, the location inside each patch where the short-space variance is locally maximum is also displayed by a black circle or a grey square for acceleration and deceleration events, respectively. (Bottom) Instantaneous solenoidal streamwise velocity fluctuation along the horizontal line indicated in the top plot. The black circle is the same point as in the top plot. The short-space variance [equation~\eqref{Eq:shortspacevar}] is also shown using a grey dashed line.}
\label{visasteps}
\end{figure}

Having computed the short-space variance at the reference plane, a condition variable C is set to non-zero values in regions where the variance exceeds the threshold, and zero otherwise. The assigned non-zero value is 1 for acceleration events and -1 for deceleration events. Mathematically, this is written as
\begin{equation}\label{Cond}
\mathrm{C}(x,z,t) = \begin{cases}1, & \text { for } \operatorname{var}>K [u^s_{rms}(y_{ref})]^2 \text { and } \dfrac{\partial {u^s}^{\prime\prime}}{\partial x}<0 \\\\
-1, & \text { for } \operatorname{var}>K [u^s_{rms}(y_{ref})]^2  \text { and } \dfrac{\partial {u^s}^{\prime\prime}}{\partial x}>0\\\\ 0, & \text { otherwise, }\end{cases}
\end{equation}
where $\partial {u^s}^{\prime\prime}/{\partial x}<0$ implies $\partial {u^s}^{\prime\prime}/{\partial t}>0$ and vice-versa, as per Taylor's hypothesis. 
This will result in patches on the reference $x-z$ plane with values of 1 and -1 as shown in figure~\ref{visasteps}. Within these patches, the location where the short-space variance is locally maximum is also shown. Let the coordinates of these locations be denoted by $(x_o,z_o)$. These coordinates, detected at $y^*\approx15$, will form the basis around which conditional averaging is performed at all wall-normal locations.

With the detected VISA locations, the conditional average of any variable $\Psi$ is then given as:
\begin{equation}\label{Eq:condavg}
    \left<\Psi\right>(\xi,y,\zeta) = \frac{1}{N}\sum_{f=1}^{N_f}\sum_{n=1}^{N_e} \Psi(x^n_o+\xi,y,z_o^n+\zeta,t^f), 
\end{equation}
where $\xi$ and $\zeta$ are the streamwise and spanwise displacements with respect to the reference or detected locations $(x_o,z_o)$, and they vary from $-L_x/2$ to $L_x/2$ and $-L_z/2$ to $L_z/2$, respectively. The inner sum is over the number of detected events ($N_e$) in a particular snapshot $f$ (at time instant $t^f$), whereas the outer sum is over the number of snapshots ($N_f$), such that the global sum of the detected events over all the snapshots is $N$. 

Note that equation~\eqref{Eq:condavg} leads to a conditional average from which phase jitter is yet to be removed \citep{johansson1987generation}. The concept of phase-jitter is explained with an example as follows. It is known that an acceleration VISA event detected at the location ($x_o, y^+\approx15, z_o$) corresponds to a wall pressure peak directly underneath, i.e. at ($x_o,y^+\approx 0,z_o$). However, there can be a small and random phase lag or lead. This means that in reality, the pressure peak may occur at a location that is randomly shifted in the streamwise-spanwise direction with respect to the detected location, i.e. it may occur at ($x_o+\Delta_x,y^+\approx 0,z_o+\Delta_z$).
This misalignment leads to a reduction in the magnitude of the pressure peak obtained after conditional averaging.

To fix this issue, we employ a cross-correlation technique \citep{johansson1987generation} that is described using the above example as follows.
We first compute the conditional average of wall pressure as usual without fixing the phase-jitter issue. We then cross-correlate this conditionally averaged wall pressure plane with the instantaneous wall pressure plane using the Fourier transform. 
Having done this, we should obtain a $x-z$ plane of correlation coefficients on the wall that displays a local maximum close to  but not necessarily at 
the point of detection, i.e. $(x_o,z_o)$. This maximum implies that the conditionally averaged wall pressure profile has its imprint in the instantaneous plane around the detection location.  
The shift between the detection location $(x_o,z_o)$ and the local maximum around $(x_o,z_o)$ gives the amount of phase lag or lead in the streamwise and spanwise directions, i.e. $\Delta_x$ and $\Delta_y$ discussed above. In order to remove the phase lag or lead, we compute a new conditional average by shifting the instantaneous planes by this
$\Delta_x$ and $\Delta_y$ around the detection points, thereby aligning them. 
Mathematically, equation~\eqref{Eq:condavg} is modified for wall pressure as
\begin{equation}\label{visafirstiter}
    \left<p^{\prime}\right>(\xi,0,\zeta) = \frac{1}{N}\sum_{f=1}^{N_f}\sum_{n=1}^{N_e} p^{\prime}(x^n_o+\Delta_x^n+\xi,\,0,\,z_o^n+\Delta_z^n+\zeta,t^f). 
\end{equation}

Now, the same procedure described for pressure at the wall can be repeated for pressure at any wall-normal location. Doing this results in $\Delta_x$ and $\Delta_y$ that depend on $y$ for each detected event. With this, equation~\eqref{visafirstiter} can be rewritten for the entire pressure field as 
\begin{equation}\label{visafirstiter2}
    \left<p^{\prime}\right>(\xi,y,\zeta) = \frac{1}{N}\sum_{f=1}^{N_f}\sum_{n=1}^{N_e} p^{\prime}(x^n_o+\Delta_x^n(y)+\xi,\,y,\,z_o^n+\Delta_z^n(y)+\zeta,t^f). 
\end{equation}
Although this gives more control on the alignment of three-dimensional conditionally averaged structures, it may result in a conditionally averaged profile that may not be very smooth in the wall-normal direction, such that layering is observed (some layering can be seen in figure~\ref{Fig:VISA}).

In the phase-jitter removal procedure, events for which the required shift is greater than approximately 40 viscous lengths in the streamwise or spanwise directions are excluded from the averaging procedure, and the total number of detected events ($N$) is reduced accordingly. Since the applied shifts are wall-normal dependent, the excluded number of events would also be wall-normal dependent.

\begin{figure}
	\centering	\includegraphics[width=1\textwidth]{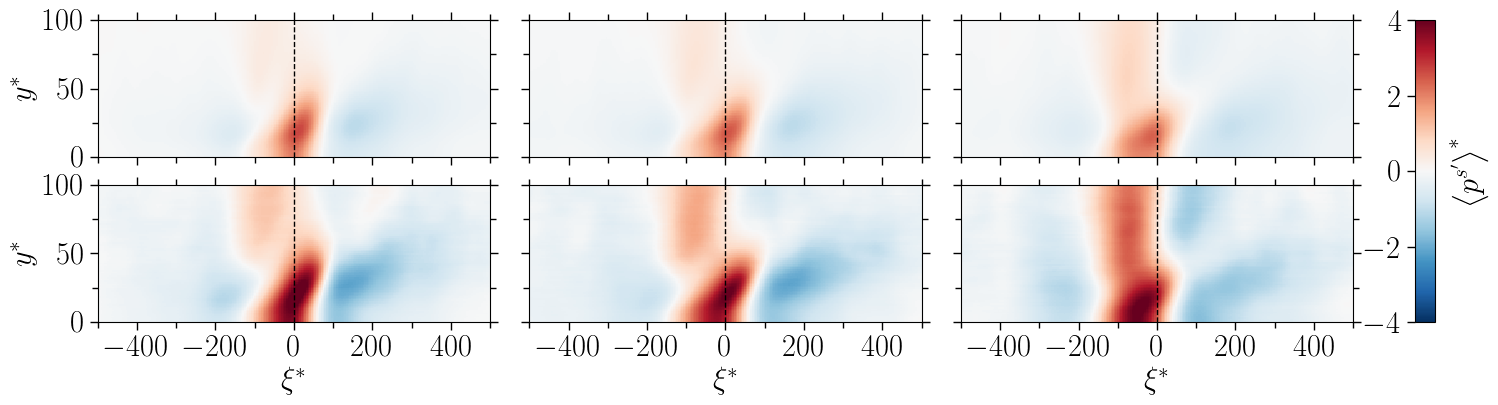}
	\caption{Contours of the solenoidal pressure along the $\xi^*-y^*$ plane at $\zeta^*=0$ after (top row) equation~\eqref{Eq:condavg} (no alignment), and after (bottom row) equation~\eqref{visafirstiter2} (first alignment iteration). The left, middle and right columns correspond to the Mach 2.28, 3 and 4 cases in table~\ref{casetab}, respectively.}
\label{Fig:Visaiters}
\end{figure}

Figure~\ref{Fig:Visaiters} shows the $\xi^*-y^*$ pressure contours taken at the centre of the shear layer, i.e. at $\zeta^*=0$, after no alignment (equation~\eqref{Eq:condavg}) and after one iteration of alignment (equation~\eqref{visafirstiter2}). As seen, the pressure contours remain qualitatively similar in both the cases, however, the magnitude after one iteration of alignment has increased substantially.

The conditionally averaged profile obtained from equation~\eqref{visafirstiter2} can be cross-correlated again with the instantaneous field, and the procedure above can be repeated to further improve the alignment. However, as noted in \cite{johansson1987generation}, and also verified for our cases, the maximum jitter is eliminated in the first iteration. Thus, the results presented in the main text are obtained after one iteration.

\section*{Acknowledgments}
This work was supported by the European Research Council grant no.~ERC-2019-CoG-864660, Critical; and the Air Force Office of Scientific Research under grants FA9550-23-1-0228 and FA8655-23-1-7016. The authors acknowledge the use of computational resources of the Dutch National Supercomputer Snellius (grant no.~2022/ENW/01251049), and of the DelftBlue supercomputer, provided by the Delft High Performance Computing Centre.

\backsection[Declaration of Interests]{The authors report no conflict of interest.}



\end{document}